\documentclass[useAMS,usenatbib]{mn2e}
\usepackage{graphicx}
\usepackage{subfigure}
\usepackage{amssymb,amsmath}
\usepackage{multirow}
\usepackage{epstopdf}

\title[AGN and supernova feedback in simulations of disc galaxies]{A study of AGN and supernova feedback in simulations of isolated and merging disc galaxies}
\author[Richard D. A. Newton and Scott T. Kay]{Richard D. A. Newton\thanks{E-mail:
newton@jb.man.ac.uk} and Scott T. Kay\\
Jodrell Bank Centre for Astrophysics, School of Physics and Astronomy, The University of Manchester, Manchester M13 9PL}
\begin{document}

\date{Accepted 2013 XXXXX XX. Received 2013 XXXXX XX; in original form 2013 XXXXX XX}

\pagerange{\pageref{firstpage}--\pageref{lastpage}} \pubyear{2013}

\maketitle

\label{firstpage}

\begin{abstract}
We perform high resolution {\it N}-body+SPH simulations of isolated Milky-Way-like galaxies and major mergers between them, 
to investigate the effect of feedback from both an active galactic nucleus (AGN) and supernovae on the galaxy's evolution. Several AGN 
methods from the literature are used independently and in conjunction with supernova feedback to isolate the most important 
factors of these feedback processes. We find that in isolated galaxies, supernovae dominate the 
suppression of star formation but the star formation rate is unaffected by the presence of an AGN. In mergers the 
converse is true when models with strong AGN feedback are considered, shutting off star formation before a starburst can 
occur. AGN and supernovae simulated together suppress star formation only slightly more than if they acted independently. 
This low-level interaction between the feedback processes is due to AGN feedback maintaining the temperature of a hot halo of 
gas formed by supernovae. For each of the feedback processes the heating temperature is the dominant parameter rather than the 
overall energy budget or timing of heating events. Finally, we find that the black hole mass is highly resolution dependent, 
with more massive black holes found in lower resolution simulations.
\end{abstract}


\section{Introduction}

In recent years, it has been recognised that the feedback energy associated with black hole accretion 
has an important influence on structure formation, on both galactic and extragalactic scales (e.g. \citealt{mcnamara2007}). 
Active galactic nuclei (AGN) are now observed as far back as $z\sim7$ (e.g. \citealt{fan2001, kurk2007, mortlock2012}) 
and it is now thought that most, if not all galaxies host a super-massive black hole at their centre. Moreover, in systems
where black hole mass can be estimated, it has been shown to correlate with the host galaxy's spheroid velocity dispersion 
(e.g. \citealt{ferrarese2000,tremaine2002}), luminosity and stellar mass (e.g. 
\citealt{magorrian1998,haringrix2004, bennert2011,mcconnell2013}). These results point to a fundamental link between the formation 
and evolution of galaxies and their central black holes; a position supported by theoretical arguments (e.g. 
\citealt{silk&rees1998,king2003,ishibashi2012}). Further evidence for the importance of AGN in the evolution of galaxies is 
provided by semi-analytic modelling of galaxy formation (e.g. \citealt{benson2003,bower2006,croton2006,guo2011}) where AGN 
feedback is invoked to reproduce the sharp drop-off in number density at the bright end of the galaxy luminosity function 
(see also \citealt{binney2004}).

The modelling of AGN in galaxy formation simulations set out by \cite{SDH2005modelling} and \cite{DMS2005,dimatteo2008}, showed 
for the first time that black hole self regulation can be achieved and the observed $M_{\rm BH}-M_{\rm bulge}$ relation reproduced 
within a fully cosmological simulation. This method is now widely employed in simulations (e.g. 
\citealt{robertson2006, dimatteo2008,khalatyan2008, johansson2009,fabjan2010,degraf2012}), although is by no means the only 
implementation currently in use and many aspects of the model are questioned. One notable assumption is that the accretion on 
to the black hole is given by the Eddington-limited Bondi-Hoyle-Lyttleton (\citealt{hoylelyttleton1939, bondihoyle1944, bondi1952}; `Bondi') 
formula which assumes spherically symmetric accretion from gas which is stationary at infinity. This has been challenged by a 
number of authors, for example \cite{boothschaye2009} dispute the correction required for under-resolved accretion flow and 
central densities around the black hole. Other authors further dispute the Bondi approach due to the inherent assumptions of 
spherical symmetry and negligible angular momentum (e.g. \citealt{debuhr2010,power2011}). \cite{hobbs2012} argue that a proper 
accretion rate estimate should take into account the gravitational potential of all matter, not just the black hole. 
Each of these modifications, as well as others taking into account more complex physics such as black hole spin 
(\citealt{fanidakis2011}), will yield different accretion rates onto the black hole and thus alter its final mass. 
Furthermore, the method by which AGN feedback energy is deposited in the surrounding gas is often varied with some works 
arguing for thermal feedback (e.g. \citealt{DMS2005,boothschaye2009,sijacki2006}) and others for kinetic or momentum feedback 
(e.g. \citealt{debuhr2010,power2011,dubois2012}) on various physical and numerical grounds. A key criticism of thermal 
feedback is that much of the energy used in raising the gas temperature is lost quickly through radiative cooling, a problem 
avoided in the method of \cite{boothschaye2009} which assumes that energy is stored until a gas particle can be heated to a 
high temperature.

Another issue that is now receiving attention is the interaction between feedback from AGN and supernovae. Typically, these 
two feedback mechanisms are treated separately in semi-analytic models (e.g. \citealt{bower2006,croton2006,guo2011}). However, 
recent work by \cite{boothschaye2013} using cosmological hydrodynamical simulations, casts doubt on the validity of this 
assumption. Their findings indicate that the combination of the two feedback processes has a weaker effect on the star formation 
rate than expected if they were acting independently. Due to the uncertainty in our present modelling of AGN feedback, it may 
be premature to assume that this interplay between AGN and supernovae is a physical effect and is not merely an outcome of 
the specific combination of models employed by \cite{boothschaye2013}.

In this paper, we seek to elucidate this interplay by studying how AGN and supernova feedback (both independently and combined) 
alter the star formation rate within simulations of isolated and merging disc galaxies. Performing such idealised simulations is 
a commonly employed method when studying such numerically demanding processes, allowing results to be produced with higher 
resolution than is normally possible in full cosmological simulations. Such systems have been used in the past to simulate a 
range of topics including galactic dynamics (e.g. \citealt{springwhite1999}), star formation (e.g. \citealt{schaye2008}), 
supernova feedback (e.g. \citealt{vecchiaschaye2008,vecchiaschaye2012}) and AGN implementations (e.g. 
\citealt{DMS2005,debuhr2010}). Our main results focus on three methods used in the literature for implementing black hole growth 
and AGN feedback, including a novel adaptation of the method suggested by \cite{power2011}. To simplify the interpretation of 
our results we adopt the same models for all other {\it sub-grid} physics included within our simulations (including supernova 
feedback, star formation and interstellar gas).

The remainder of the paper is organised as follows. In Section \ref{modelTheory} we describe our numerical method, namely how we 
set up our initial conditions and give details of the various sub-grid implementations, including the supernova and AGN feedback 
models. Our main results are presented for isolated galaxies in Section \ref{isoGalaxies} and merging galaxies in Section 
\ref{mergeGalaxies}, while in Section \ref{discussion} we take a closer look at some of our model assumptions. Finally, conclusions 
are drawn and future work outlined, in Section \ref{conclusions}.

\section[]{Numerical Method}
\label{modelTheory}

\begin{table*}
\caption{Numerical parameters for our model galaxies, simulated with varying resolution. Columns 2-6 give the total particle 
number, as well as the number in the DM halo, gaseous disc, stellar disc and stellar bulge respectively. Columns 7 \& 8 list the 
dark matter and baryonic particle masses while Column 9 gives the (equivalent Plummer) gravitational softening length used in our 
simulations. Values in parentheses correspond to a version of the low resolution run with modified softening (MS). All of our main results use the high resolution simulations; runs using the other resolution levels are investigated in Section \ref{res_effects}.}
\begin{center}
\begin{tabular}{p{15mm} c c c c c c c l}
\hline
Resolution & $N$ & $N_{\rm h}$ &  $N_{\rm g}$ &  $N_{\star}$ &  $N_{\rm b}$ &  $m_{\rm DM}\,[{\rm M_{\odot}}]$ & $m_{\rm baryon}\,[{\rm M_{\odot}}]$ & $\varepsilon_{\rm soft}\,[{\rm kpc}]$ \\
\hline
Very-Low  & $7\times10^4$  & $\phantom{00}59109$ &  $\phantom{00}2178$ &  $\phantom{00}6099$ &  $\phantom{00}2614$ & $3.08\times10 ^7$ &  $6.24\times 10^6$ & $0.23$ \\
Low(MS)  & $7\times10^5$  & $\phantom{0}591089$ &  $\phantom{0}21780$ &  $\phantom{0}60990$ &  $\phantom{0}26138$ & $3.08\times10 ^6$ &  $6.24\times 10^5$ & $0.05$($0.11$) \\
High & $7\times10^6$  & $5910890$ &  $217800$ &  $609900$ &  $261380$ & $3.08\times 10^5$ &  $6.24\times 10^4$ & $0.05$ \\
\hline
\end{tabular}
\end{center}
\label{tab:Resparams}
\end{table*}

All simulations discussed in this paper are performed using a modified version of the publicly-available 
${\it N}$-body + Smoothed Particle Hydrodynamics (SPH; see e.g. \citealt{monaghan1992,springel2010}) code {\sc Gadget-2} 
(\citealt{gadget2}). We first outline our method for setting up idealised disc galaxies, before discussing the sub-grid physics 
used in the simulations.

\subsection{Initial conditions}
\label{ICsection}
The full method for constructing equilibrium disc galaxies is given in Appendix \ref{ICappendix}; here we provide a summary. We 
largely follow \cite{SDH2005modelling} which builds on the earlier work of \cite{MMW}. Our approach differs in a few key areas 
such as our treatment of adiabatic contraction of the dark matter (DM) halo (see Section \ref{Hcontract}) and the model used to 
describe the interstellar medium (ISM; Section \ref{radCoolingISMSect}). Our model galaxies consist of a DM halo surrounding a 
disc made up of stars and gas, with a central stellar bulge. We assume that the DM halo and bulge follow Hernquist profiles and 
that the disc components are axisymmetric, following an exponential surface density profile with both gas and stellar components 
having the same scale length. The gas disc is assumed to be in hydrostatic equilibrium, whereas the stellar component is an 
isothermal sheet. 

Our galaxy parameters are as follows. The total mass is $M_{200}=1.36\times10^{12}\, {\rm M_\odot}$\footnote{$M_{200}$ is the mass 
enclosed within $R_{200}$, the radius at which the mean density is $200 \rho_{\rm cr}$, where $\rho_{\rm cr}$ is the critical matter 
density required for a flat universe.}, with a Navarro, Frenk \& White (NFW; \citealt{NFW1997}) concentration parameter, $c=9$. 
The mass is split (stated as fractions of $M_{200}$) according to the disc fraction $f_{\rm d}=0.04$ and bulge fraction 
$f_{\rm b}=0.01$, with the remaining mass residing in the DM halo. (We do not include a gaseous halo in this study but note 
that one quickly forms when supernova feedback is included.) The disc is split once more into gas and stars with the mass 
fraction of the disc in gas, $f_{\rm g}=0.3$. We set the radial disc scale length, $h=3.45\,{\rm kpc}$, through angular momentum 
considerations and use it to define the bulge scale length and disc scale height $b=z_0=0.2\, h$. Our galaxy parameters are chosen 
to approximately describe a Milky-Way-like galaxy and allow comparison with earlier work such as \cite{SDH2005modelling,schaye2008} 
and \cite{debuhr2010}.

The galaxies were constructed in both high and low resolution realisations and were run for $3\,{\rm Gyr}$. Table 
\ref{tab:Resparams} summarises our choice of numerical parameters. The additional physical processes affecting the gas (over and 
above gravity and standard SPH) are discussed below.

\subsection{Radiative cooling, ISM \& star formation}
\label{radCoolingISMSect}

Gas is allowed to cool radiatively following the method described by \cite{kay2004}, who adopted the approach of \cite{thomas1992}. 
We utilise a tabulated equilibrium cooling function from \cite{sutherland1993} for a metallicity $Z = 0.3\, {\rm Z_\odot}$ gas.
The temperature of the gas is not permitted to fall below a critical value $T_{\rm c}=10^4\, {\rm K}$, at which point 
we assume the gas temperature is maintained by the meta-galactic UV background. 

The ISM and star formation model employed is that of \cite{schaye2008}.
We assume that the ISM is an ideal multi-phase gas which may be described by a 
polytropic equation of state above a critical value for the number density of hydrogen, $n_{\rm H,c}$
\begin{equation}
\label{EoS}
P = \left\{ \begin{array}{ll}
An_{\rm H}^{\gamma_{\rm eff}} & \mbox{ if $n_{\rm H}>n_{{\rm H,c}}$,} \\
 n_{\rm H} k_{\rm B} T_{\rm c} / ( X\mu) & \mbox{ otherwise,}
\end{array} \right.
\end{equation}
where $A$ is a constant set by ensuring the pressure is continuous across the boundary $n_{\rm H}=n_{\rm H,c}$
and $\gamma_{\rm eff}=4/3$. The critical density is set to $n_{\rm H,c}=0.1 {\rm cm^{-3}}$, the mean molecular weight, $\mu=0.59$ 
and the hydrogen mass fraction, $X=0.76$, the latter two having primordial values. Gas may leave the equation 
of state when it receives (AGN or supernova) feedback energy or if its thermal energy increases 
by $0.5\, {\rm dex}$ in a single timestep (as in \citealt{boothschaye2009}). Allowing particles to leave the equation of state in this fashion 
prevents artificial damping of feedback driven outflows.

\begin{figure}
\begin{center}
\includegraphics[width=\columnwidth]{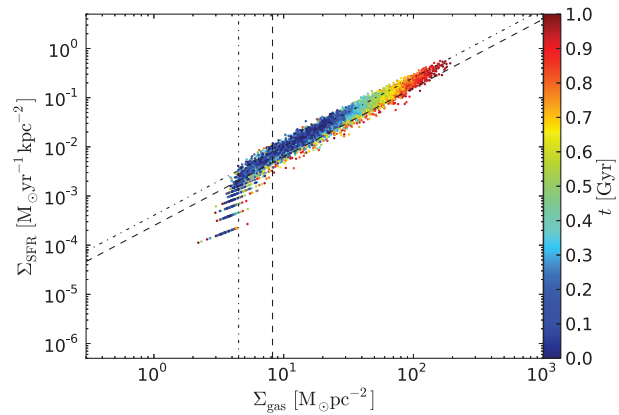}
\end{center}
\caption{Recovered Kennicutt-Schmidt relation for a high resolution disc galaxy evolved for a time, $t=1\,{\rm Gyr}$. Each point represents
the star formation rate density averaged within an annulus centred on the galaxy rotation axis, while the colour represents the age of these stars 
at the end of the simulation (with red points being the oldest stars). The dashed lines show the input Kennicutt-Schmidt relation and density 
cut-off, while dot-dashed lines show these adjusted for $f_{\rm g}=0.3$ (the value used in our initial conditions).}
\label{fig:KenSchmidt}
\end{figure}

A gas particle lying on the equation of state may be converted to a collisionless star particle representing a simple 
stellar population (SSP) described by a Salpeter initial mass function (IMF). The volumetric star formation rate is designed to match the observed Kennicutt-Schmidt law 
(\citealt{kennicutt1998}) for a galaxy with disc gas fraction, $f_{\rm g}=1$ (the commonly-employed value in cosmological simulations). Specifically, the law numerically reproduced here is
\begin{equation}
\dot{\Sigma}_\star=2.5\times 10^{-4} {\rm M_\odot yr^{-1} kpc^{-2}} \left( \frac{\Sigma_{\rm g}}{1\, {\rm M_\odot pc^{-2}}} \right)^{1.4},
\end{equation}
where $\dot{\Sigma}_\star$ is the star formation rate surface density and $\Sigma_{\rm g}$ is the gas surface density. 
Fig. \ref{fig:KenSchmidt} demonstrates the good match between simulated and observed star formation rates, once the
latter is adjusted for the disc gas fraction adopted for our initial conditions ($f_{\rm g}=0.3$, lower than the value assumed in deriving the volumetric star formation rate; see \citealt{schaye2008} for details)
\footnote{We have also checked that our implementation reproduces the star formation rates found by \cite{schaye2008}
when we adopt a zero metallicity gas, for both Salpeter and Chabrier IMFs.}.

\subsection{Supernova feedback}
\label{SNfeedbackmethod}
The supernova feedback implementation described here is motivated by the desire to ensure high temperature gas is produced, as 
this is more likely to have an impact on the star formation rate. This is because gas with $T\gg 10^6\, {\rm K}$ cools more slowly than 
cooler gas, mainly via thermal bremsstrahlung, allowing the heated material to interact hydrodynamically with the surrounding gas and 
escape the star forming region. We guarantee a high temperature by reducing the mass loading factor of the heated gas, given a fixed amount of energy
available from the supernovae (see \citealt{vecchiaschaye2012} for an application of this method to disc galaxies and \citealt{kay2003} 
for galaxy groups). 

We assume that a fraction of the stars in each SSP will end their lives as type II supernovae (SNII), releasing energy to the environment. This 
typically occurs on a short timescale which we assume to be negligible. 
The amount of energy provided by SNII per unit mass of stars formed is calculated as
\begin{equation}
\begin{split}
u_{\rm SN} & = E_{\rm SN}\,n_{\rm SNII}, \\
& = E_{\rm SN}\int_{M_0}^{M_1}\Phi(M)\,{\rm d}M, \\
& = 5.6 \times 10^{15}\,{\rm erg\,g^{-1}} \left(\frac{E_{\rm SN}}{10^{51}{\rm erg}} \right ) \left( \frac{n_{\rm SNII}}{1.11 \times 10^{-2}{\rm M}_\odot^{-1}}\right ),
\end{split}
\end{equation}
where $E_{\rm SN}$ is the energy per Type II supernova and $n_{\rm SNII}$ is the number of supernovae per unit mass, obtained by integrating over the IMF ($\Phi$) 
between $M_0=0.1\,{\rm M_\odot}$ and $M_1=100\,{\rm M_\odot}$. The specific energy change imposed 
on the local gas is then
\begin{equation}
u_{\rm heat}=\epsilon_{\rm SN} \left[\frac{m_{\rm star}}{m_{\rm gas}N_{\rm heat}} \right ] u_{\rm SN},
\end{equation}
where $\epsilon_{\rm SN}$ is a dimensionless supernova efficiency parameter (accounting for
small-scale energy losses) and the term in brackets is the inverse mass loading
factor (heating $N_{\rm heat}$ gas particles of mass $m_{\rm gas}$ for every new star particle with mass $m_{\rm star}$). 
The heating temperature can then be expressed as
\begin{equation}
\begin{split}
T_{\rm heat} \approx 2.7 & \times 10^7\, {\rm K} \left(\frac{\epsilon_{\rm SN}}{N_{\rm heat}} \right ) \left(\frac{m_{\rm star}}{m_{\rm gas}} \right ) \\
& \left(\frac{E_{\rm SN}}{10^{51}{\rm erg}} \right ) \left( \frac{\mu}{0.59}\right ) \left( \frac{n_{\rm SNII}}{1.11 \times 10^{-2}{\rm M}_\odot^{-1}}\right ). 
\end{split}
\end{equation}
For the purpose of our fiducial simulations we set $T_{\rm heat} = 10^7\, {\rm K}$ and $N_{\rm heat}=1$, yielding an assumed 
efficiency of $\epsilon_{\rm SN}\approx 0.4$ (this choice is investigated further in Section \ref{snTemp}).
When heated by supernovae, a gas particle is removed from the equation of state on its current timestep and tagged as a wind particle. However, 
should the particle meet the conditions for joining the equation of state on any subsequent timestep, it will return as normal. 

\subsection{Black hole growth and AGN feedback}
\label{AGN_models}
There exists a large variety of work employing different methods for simulating AGN and their effects on the host galaxy. In this paper we study three of
the most widely used implementations, namely those by \cite{SDH2005modelling,boothschaye2009}; and 
a novel method based on the model of \cite{power2011}. The AGN models may be simplistically broken down into three processes: formation, growth and feedback. 
The formation of black holes at high redshift is an interesting topic in itself and an active area of research. However, it is beyond 
the scope of this paper where we simply assume our galaxies already contain a super-massive BH at their centre. We discuss the other two processes in turn.

\subsubsection{Black hole growth: the Bondi method}
\label{AccretionSect}
\cite{SDH2005modelling} and \cite{boothschaye2009} both used modified versions of the Bondi-Hoyle-Lyttleton 
(\citealt{hoylelyttleton1939, bondihoyle1944, bondi1952}) model to calculate the accretion rate 
onto the black hole from a surrounding gas reservoir, given by
\begin{equation}
  \label{BHLeqn}
  \dot{M}_{\rm BH}=\alpha\frac{4\pi G^2 M^2_{\rm BH}\rho_{\rm gas}}{(c_{\rm s}^2+v^2)^{3/2}},
\end{equation}
where $\rho_{\rm gas}$ is the density of the surrounding gas, $c_{\rm s}$ its sound speed, and $v$ the velocity of the 
BH relative to the gas. The additional factor $\alpha$ was added by \cite{SDH2005modelling} to counteract an assumed underestimate for the accretion rate due to inadequate numerical resolution. 
Simulating the accretion fully requires proper resolution of the Bondi radius
\begin{equation}
 r_{\rm B}\approx \frac{GM_{\rm BH}}{c_{\rm s}^2},
\end{equation}
which is considerably smaller than the force resolution in current simulations, for the case of galactic-scale black holes. 
Furthermore, one must also resolve the multiphase ISM and crucially its cold dense clouds, to allow accurate computation 
of the gas density and sound speed close to the BH.

\cite{boothschaye2009} modified the correction factor, $\alpha$, arguing that the accretion is 
not always equally under-resolved and suggested the following scaling with density 
\begin{equation}
\alpha = \left\{ \begin{array}{cl}
 1 &\mbox{ if $n_{\rm H}<n_{\rm H,c}$} \\
 \left( \frac{n_{\rm H}}{n_{\rm H,c}}\right)^\beta &\mbox{ otherwise,}
       \end{array} \right.
\end{equation}
where $n_{\rm H,c}$ is the same critical value as for the ISM and star formation ($n_{\rm H,c}=0.1\, {\rm cm^{-3}}$). 
The impact of varying the power-law index, $\beta$, was investigated by \cite{boothschaye2009} as well as varying the constant $\alpha$ value, as used in the more standard 
formalism. They found that observables are much more sensitive to the choice of $\alpha$ than to any reasonable change in $\beta$. We have implemented both
methods as laid out in their original form.

A BH particle's mass is tracked by two variables: the {\it dynamical} mass which increases instantaneously when a particle is captured by the BH; and the smoothly integrated 
{\it internal} mass which is used to determine the accretion rate of the BH. In practice, the BH particle grows by stochastically capturing surrounding gas particles within the smoothing length at a rate which coarsely matches the internal black hole mass (see e.g. \citealt{DMS2005, sijacki2007, dimatteo2008, boothschaye2009}). 
Upon accreting a particle the mass is added to that of the black hole. 
The potential drawback of using the SPH kernel is that gas particles may be captured at arbitrarily large distances 
from the sink particle due to the use of an adaptive smoothing length (discussed e.g. by \citealt{debuhr2010, power2011}).

\subsubsection{Black hole growth: the accretion disc method}

The \cite{power2011} method does not use the Bondi-Hoyle-Lyttleton equation. Instead, it assumes that any particle which 
passes within a fixed accretion radius, $r_{\rm acc}$, is captured onto the {\it accretion disc} surrounding the BH. This mass is then
moved to the BH on a viscous timescale.
A key feature of this method is that, unlike the Bondi-Hoyle-Lyttleton accretion, only gas with low angular momentum is likely to be captured.
In simulations lacking sufficient spatial resolution to simulate true accretion discs existing on sub-${\rm pc}$ scales (e.g. \citealt{kato1998}), it is favourable to set 
$r_{\rm acc}$  to the smallest resolvable length-scale, 
i.e. the gravitational softening length $\varepsilon_{\rm soft}$; this is the approach taken here ($\varepsilon_{\rm soft}=50\,{\rm pc}$). 

\begin{table*}
\caption{Summary of our main runs and important parameter choices. Column 1 lists the label used to refer to each model,
while columns 2 \& 3 specify whether supernovae and AGN are included. Pertinent details of the run are given in column 4; columns 5-7
lists values of the appropriate BH accretion rate parameter; column 8 the number of particles heated by the BH in one event 
and column 9 the assumed value for the feedback coupling efficiency parameter. The value of $N_{\rm heat}$ for the KAGN
runs is approximate.}
\begin{center}
\begin{tabular}{p{20mm} c c p{70mm} c c c c c}
\hline
Label & SN & AGN & Details & $\alpha$ & $\beta$ & $\epsilon_{\rm ff}$ & $N_{\rm heat}$ & $\epsilon_f$ \\
\hline
NFB & N & N & No feedback & - & - & - & - & - \\
SN & Y & N & Supernovae only & - & - & - & - & - \\
KAGN & N & Y & Bondi accretion, kernel feedback & $100$ & - & - & 50 & $0.05$ \\
KAGN\_SN & Y & Y & Bondi accretion, kernel feedback, supernovae & $100$ & - & - & 50 & $0.05$ \\
SAGN & N & Y & Boosted Bondi accretion, strong feedback & - & 2 & - & 10 & $0.15$ \\
SAGN\_SN & Y & Y & Boosted Bondi accretion, strong feedback, supernovae & - & 2 & - & 10 & $0.15$ \\
DAGN & N & Y & Disc accretion, strong feedback & - & - & 0.1 & 10 & $0.15$ \\
DAGN\_SN & Y & Y & Disc accretion, strong feedback, supernovae & - & - & 0.1 & 10 & $0.15$ \\
\hline
\end{tabular}
\end{center}
\label{tab:Simparams}
\end{table*}

Preliminary testing of this method found it to be unphysical at the typical resolution of our simulations ($\sim 100\, {\rm pc}$), yielding 
an artificially large accretion rate, an overly-massive black hole and a large cleared-out area at 
the centre of the galaxy. \cite{wurster2013} attempt to overcome this problem by significantly reducing 
$r_{\rm acc}$ to limit mass accretion, with the precise value being varied as part of their investigation.
It is, however, unphysical to take particles as point masses in this regime since capturing one whole corresponds 
to a gas cloud with radius of order the SPH smoothing length instantaneously collapsing onto the much smaller accretion disc. 
We take an alternative route by instead interpreting the accretion radius as a scale within which the captured
material is gravitationally bound to the black hole system and will contribute to the mass of the accretion disc
on a timescale set by the local free-fall time, $t_{\rm ff} = 1/ \sqrt{G \langle \rho_{\rm gas} \rangle}$. 
Specifically, the gas mass is added to the disc at a rate given by
\begin{equation}
\frac { {\rm d}M_{\rm disc}}{ {\rm d}t}  = -\frac{\epsilon_{\rm ff}M_{\rm gas}(<r_{\rm acc})}{t_{\rm ff}},
\label{discMflow}
\end{equation}
where $\epsilon_{\rm ff}$ is an efficiency parameter that may represent unresolved effects such as turbulence or 
residual angular momentum (we set $\epsilon_{\rm ff}=0.1$, as discussed in the next section). 
Motivating the large scale accretion via the freefall timescale is similar in spirit to \cite{hobbs2012}, who did so 
by modifying the Bondi model. Mass is then added to the BH at the rate
\begin{equation}
{\dot M}_{\rm BH} =  {\rm min}\left[\frac{M_{\rm disc}}{t_{\rm visc}}, {\dot M}_{\rm Edd} \right],
\end{equation}
where the viscous timescale $t_{\rm visc}$ may be estimated from physical arguments and is typically set to 
$t_{\rm visc} \sim 10-100\, {\rm Myr}$ (\citealt{kato1998,power2011}; we set $t_{\rm visc}=10\,{\rm Myr}$). For all three methods, we
restrict the accretion rate to the Eddington limit
\begin{equation}
  \dot{M}_{\rm Edd}=\frac{4\pi G M_{\rm BH}m_{\rm H}}{\epsilon_{\rm r}\sigma_{\rm T}c},
\end{equation}
where we assume the standard value for the radiative efficiency parameter, $\epsilon_{\rm r}=0.1$.

Independent of accretion rate method, particles are numerically captured similarly to \cite{SDH2005modelling}. Particles are accreted only when the 
system mass (BH and accretion disc if applicable) exceeds the BH particle's dynamical mass. Accreted particles would ideally contribute momentum to the BH particle as it is 
clearly desirable to conserve momentum in all simulations. However we found that artificially large kicks and 
two-body scattering from other particles caused the black hole to become displaced from the centre of the galaxy, even in isolated quiescent galaxies. We 
therefore apply the approach of \cite{DMS2005} and \cite{boothschaye2009} to all three methods, and re-centre the black hole particle on the neighbouring ($r_{ij}<h$) 
particle of minimum potential until it is a factor of 10 more massive than a single gas particle.

\begin{figure*}
\begin{center}
\includegraphics[width=14cm]{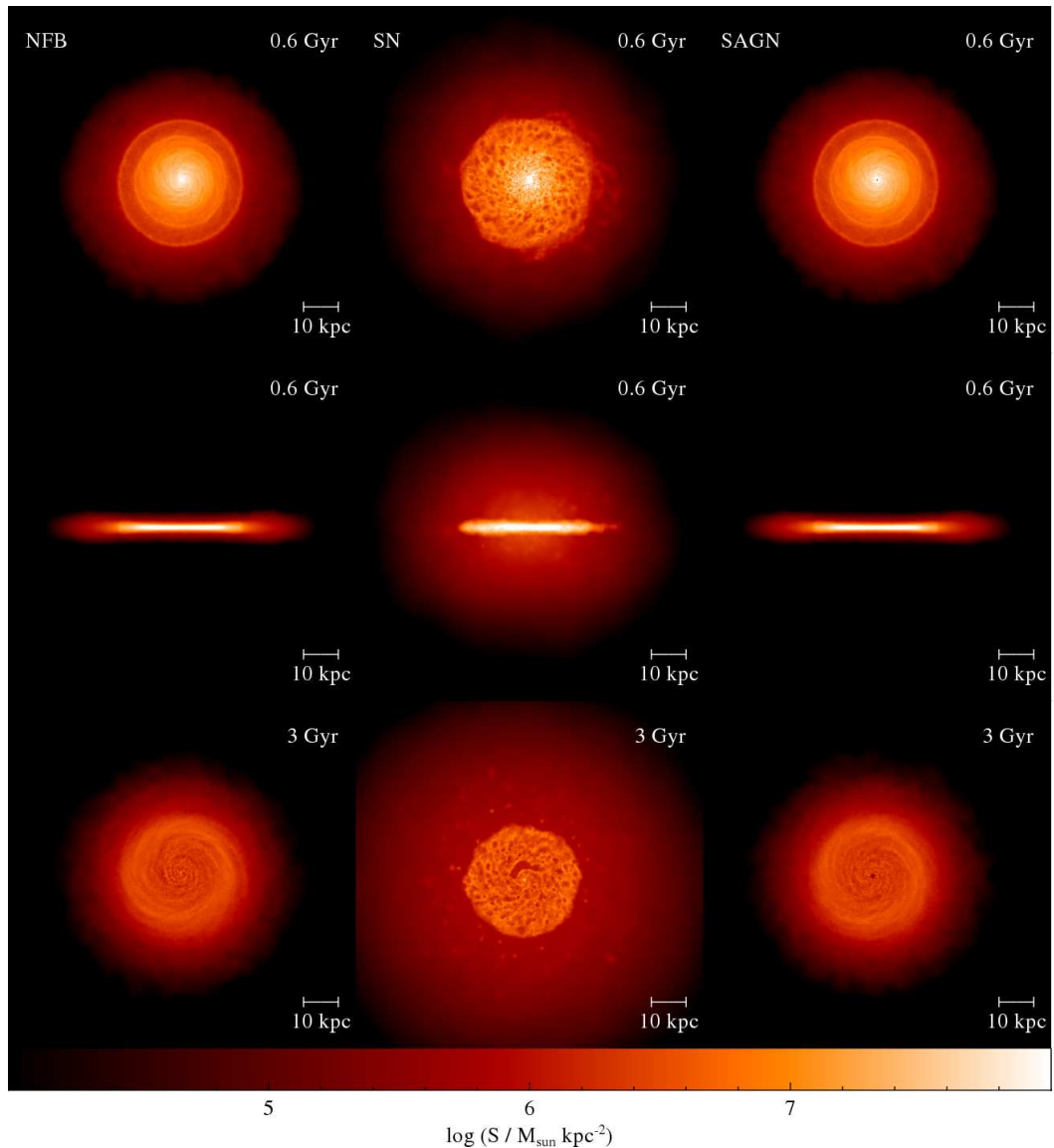}
\end{center}
\caption{Projected face- and edge-on gas surface density maps for an isolated galaxy simulated with various feedback implementations, 
at $t=\,0.6$ and $3\,{\rm Gyr}$. Columns show results for the NFB, SN and SAGN models respectively. The panels are $100\,{\rm kpc}$ on a 
side and surface density is coloured on a log scale. The reduction in density with time, seen for all models, occurs due to star formation. 
A slight redistribution of gas is seen for the SAGN model but the most pronounced case is for supernova feedback (SN), where a hot halo is 
quickly generated.}
\label{fig:DGpretty}
\end{figure*}

\subsubsection{AGN feedback}
The accretion rate onto the central BH is used to determine the energy budget for the AGN feedback. It is commonly assumed that 
a fraction of the accreted rest mass energy is released from the BH, with an additional multiplicative factor to represent the efficiency with 
which this then couples to the surrounding gas
\begin{equation}
\label{feedbacke}
{\dot E}_{\rm feed}=\epsilon_{\rm f} L_r=\epsilon_{\rm f} \epsilon_{\rm r} {\dot M}_{\rm BH} c^2,
\end{equation}
where $\epsilon_{\rm r}=0.1$ is the radiative efficiency as before and $\epsilon_{\rm f}$ is the feedback coupling efficiency.

The most accurate way to distribute this feedback energy, and the form it takes, is unclear. One method is to utilise an SPH-like approach 
and simply deposit the energy thermally in a kernel-weighted fashion; such an approach was taken by \cite{SDH2005modelling}. This method is 
numerically simple and characterises a high energy quasar radiation field coupling isotropically to the surrounding gas. 
However, releasing the feedback energy in this way may not be able to provide gas hot enough to drive strong outflows. 
An alternative approach was taken by \cite{boothschaye2009}. By `bottling up' the energy it is ensured 
that a critical temperature change, $\Delta T_{\rm min}$, is reached before a heating event occurs. The energy which must be stored 
to heat $N_{\rm heat}$ gas particles by $\Delta T_{\rm min}$ is
\begin{equation}
E_{\rm crit}=\frac{N_{\rm heat} m_{\rm gas} k_{\rm B} \Delta T_{\rm min}}{(\gamma - 1)\mu m_{\rm H}}.
\end{equation}
Once heated, it is then assumed that the gas will rise buoyantly in the ambient medium, mimicking outflows found in observations.

In the following sections we perform simulations that employ the methods for describing gas processes and feedback outlined above. 
Specifically, we simulate the Bondi accretion method with kernel-weighted feedback as set out by \cite{SDH2005modelling}, labelled here as K[ernel]AGN;
the Bondi method with `strong' feedback by \cite{boothschaye2009}, labelled S[trong]AGN; and the disc accretion method based on \cite{power2011} 
with `strong' feedback (D[isc]AGN). We also consider a model with no feedback\footnote{To clarify, NFB runs contain no macroscopic winds, however a pressure contribution from supernova feedback is implicit in the assumed EoS for all simulations.} (NFB) and a model with supernovae only (SN). Models run with both 
AGN and supernovae are denoted XAGN\_SN, where X is (K,S,D) as discussed. Table \ref{tab:Simparams} summarises the details of these runs 
and lists values for the main AGN feedback parameters.

\section[]{Isolated disc galaxies}
\label{isoGalaxies}

We begin our investigation by performing tests of the feedback recipes in isolated disc galaxies. Performing simulations of an isolated disc galaxy provides us the 
opportunity to study the models in a regime where the black hole is typically expected to be in a phase of low activity and provides a comparison for the 
behaviour of black holes in violent mergers. Employing such high resolution `toy models' also ensures any observed phenomena may be 
more easily disentangled because the systems themselves are simple and relatively well understood.

\subsection{Simulations with and without supernova feedback}
\label{SNsectDG}

We first investigate the effect of supernovae on the host galaxy properties, neglecting the AGN for the time being. 
Fig. \ref{fig:DGpretty} shows projected gas surface density maps for the NFB and SN runs at $t=\,0.6$ and $3\,{\rm Gyr}$. The maps illustrate 
qualitatively that supernovae quickly generate a gas halo that persists for the duration of the simulation. They also modify the disc structure, increasing the 
gas porosity. The reduction in density in the disc over time is predominantly due to star formation as this can be seen for the NFB case.

Fig. \ref{fig:DGsfrsx} shows the evolution of the global star formation rate for the NFB and SN simulations (solid and dashed lines respectively). We also 
show the star formation suppression factor, $S_{\rm X} = {\dot M_{\star}} ({\rm NFB}) / {\dot M_{\star}} (\rm SN)$, in the bottom panel. 
As expected, we see a reduction in the star 
formation rate when supernovae are included. This persists until $t\sim 1.5\, {\rm Gyr}$ when we see a relative {\it increase} due to low-level star formation from gas which would have otherwise already formed stars in the absence of supernovae\footnote{
The suppression in the star formation rate is found to be consistent with that seen in \cite{vecchiaschaye2012} under the appropriate parameter choices (see 
their Fig. 10).}.

The primary change seen in simulations with supernova feedback is a reduction in the stellar mass of the disc. This is clearly seen in Fig. \ref{fig:DG_sn_Mphase}, 
where both the mass in new stars and in a supernova-driven wind is shown as a fraction of initial gas mass, versus time for the NFB and SN simulations. 
A reduction of $\sim10$ per cent in stellar mass is seen by the end of the simulation, caused by the inclusion of supernovae. It is clear that the majority of supernova 
heated gas is returning to the equation of state on a short timescale, as the mass in the wind is considerably lower than in new stars.

\subsection{Simulations with AGN feedback only}

\label{IsoGalAGNfdbck}
\begin{figure}
\begin{center}
\subfigure{\includegraphics[width=\columnwidth]{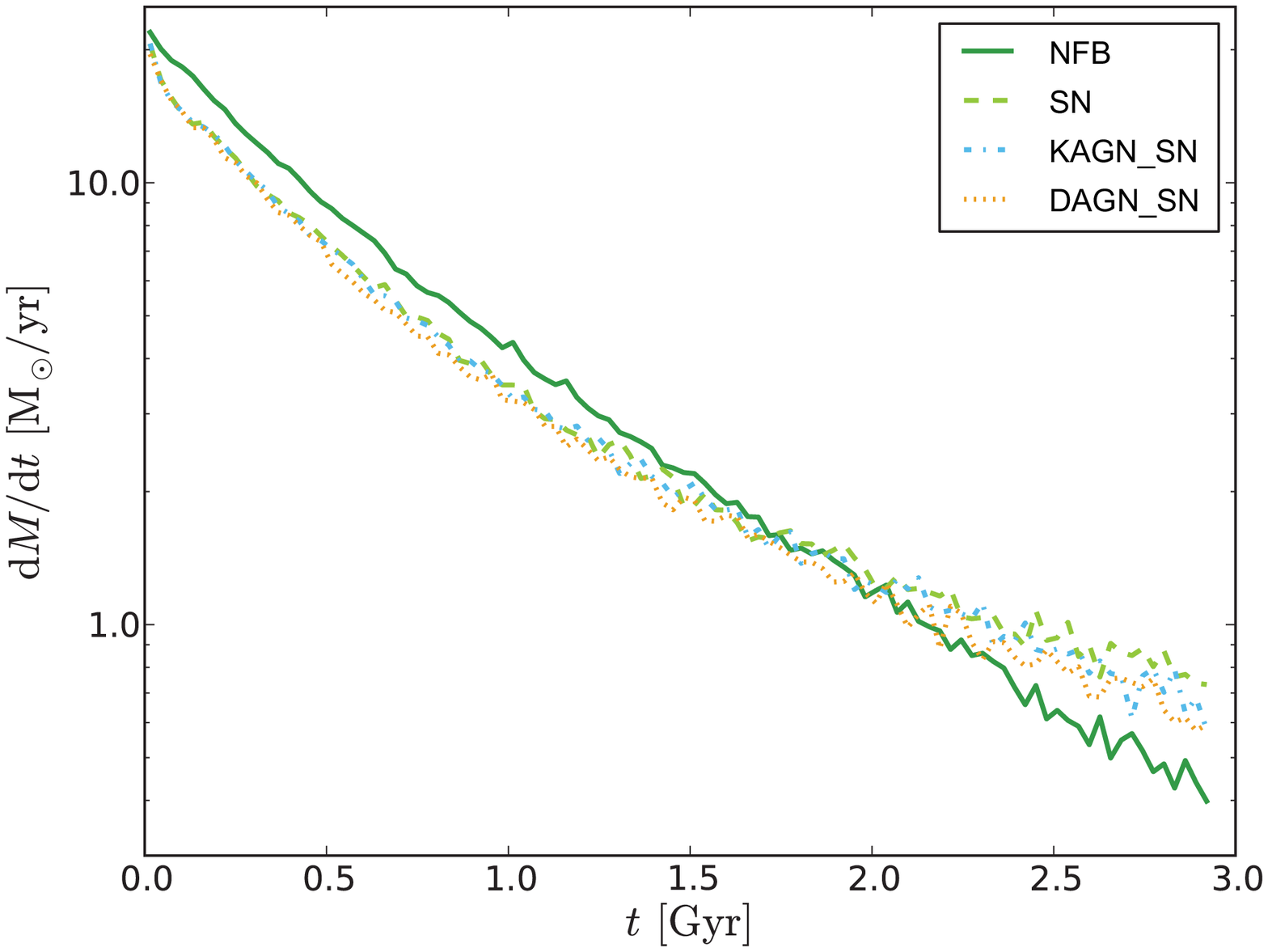}}
\subfigure{\includegraphics[width=\columnwidth]{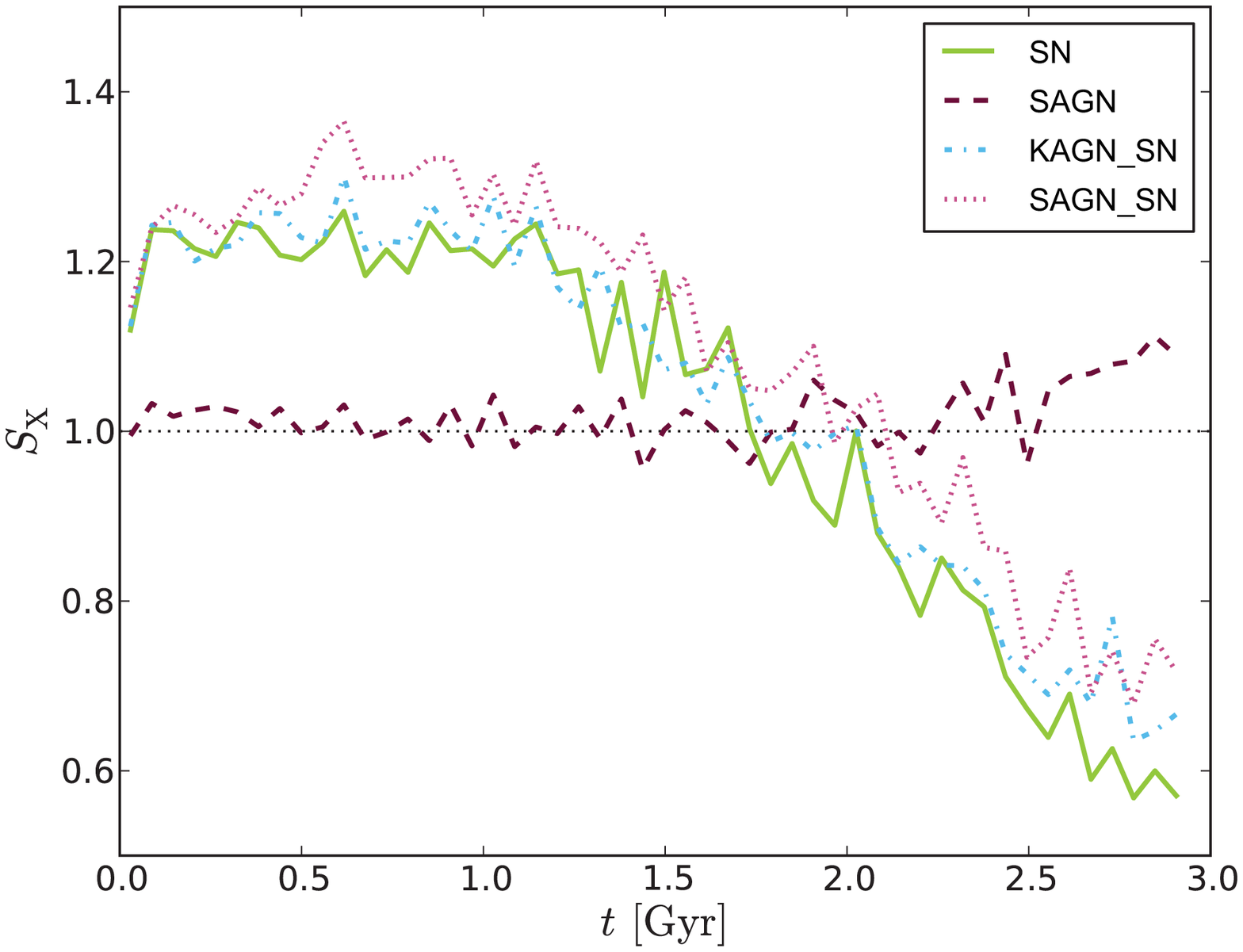}}
\end{center}
\caption{Star formation rate (top) and suppression factor (bottom) throughout simulations of an isolated high-resolution disc galaxy. Suppression 
is calculated as a ratio of instantaneous star formation rates, no-feedback case over feedback case. Supernova feedback causes the dominant suppression 
of the star formation rate, while the AGN wind is unable to interact with the majority of the galaxy, only having a minor influence when able to couple 
with the supernova-inflated hot halo.}
\label{fig:DGsfrsx}
\end{figure}

\begin{figure}
\begin{center}
\includegraphics[width=\columnwidth]{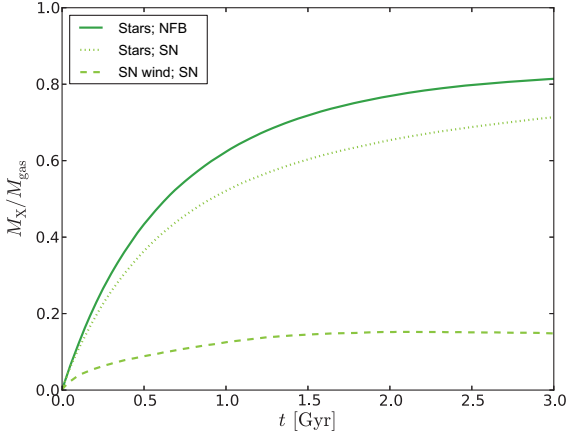}
\end{center}
\caption{Fractional mass in new stars and particles tagged as being in a supernova wind versus time for the isolated high-resolution 
disc galaxy, with and without supernova feedback. All masses are shown as a fraction of the initial gas mass. Supernovae act to reduce 
the amount of gas on the equation of state and therefore the mass in stars.}
\label{fig:DG_sn_Mphase}
\end{figure}

We now consider runs with AGN feedback only. Unlike the supernovae, the AGN feedback has no effect on the star formation rate for 
any model (an example is shown in the bottom panel of Fig.~\ref{fig:DGsfrsx} for the SAGN run). This is likely due to the differing 
scales involved as the black hole typically only interacts with the central $\sim 0.1\, {\rm kpc}$ whereas the star formation takes 
places on ${\rm kpc}$ scales. Although the black holes do not change the large-scale galaxy properties they can be seen to drive a 
weak wind.

\begin{figure}
\begin{center}
\includegraphics[width=\columnwidth]{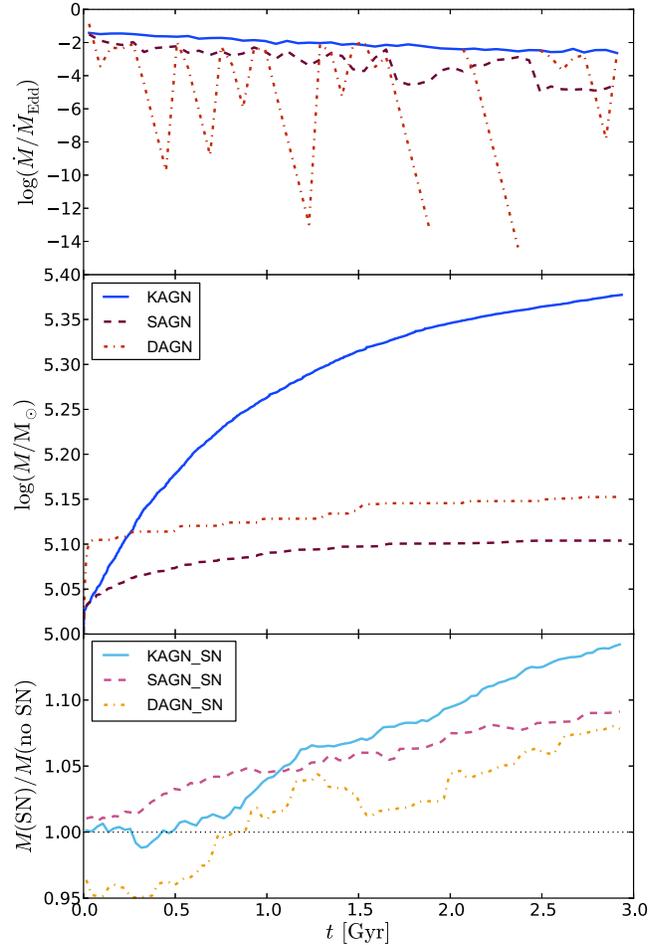}
\end{center}
\caption{Evolution of BH mass with time, for the fiducial AGN models without supernovae in isolated galaxy simulations. The top panel shows the smoothed black hole accretion rate in units of the Eddington rate, while the central panel shows the integrated black hole mass in each of the three models. The bottom panel shows ratios of integrated masses found 
between these and corresponding simulations including supernovae. Supernovae increase the BH mass in all models by preventing 
gas depletion through star formation and driving gas towards the BH. However, this increase ($\sim 10$ per cent) is much smaller than the differences 
between the AGN feedback models themselves (which can be tuned if necessary).}
\label{fig:DGbhMassEvo}
\end{figure}

The growth in BH mass with time is shown in, where we plot both the smoothed accretion rate (in units of the Eddington rate; top panel) and the integrated mass (middle panel). Here we can see that the 
three models produce different accretion behaviour, leading to very different final masses. However, we note that the normalisation and therefore the final 
mass may easily be tuned through accretion rate parameter choices for any given model. We have chosen to employ the original values 
where appropriate (KAGN and SAGN) as these were tuned to recover the $M_{\rm BH}-M_{\rm bulge}$ relation in full cosmological 
simulations. For the \cite{power2011} method (which was not tested on cosmological volumes) we adopted the original 
value for $t_{\rm visc}(=10\,{\rm Myr})$ and tuned the free-fall efficiency parameter, $\epsilon_{\rm ff}=0.1$, so as to bring the final mass roughly in 
line with the \cite{boothschaye2009} method as we adopted their feedback prescription for the DAGN model. The most significant factor 
affecting the final BH mass in isolated galaxy simulations is the resolution used (we investigate this in Section \ref{res_effects}). 

We also note that the models predict different BH histories. Bondi-based methods predict that the BH
grows smoothly whereas in the DAGN method it undergoes an initial period of rapid growth followed by quasi-periodic accretion events. 
Unlike the spatially adaptive Bondi methods, the disc method employs a small and finite accretion radius which initially contains gas (an
artefact of our initial conditions) but is quickly removed through accretion and feedback, shutting off the black hole growth. After this time,
the net BH growth rate is similar in the SAGN and DAGN models despite the different accretion method, while the BH in the KAGN method continues to grow at a faster rate. This suggests that the choice of feedback
plays an important role in determining the BH growth rate (we confirm this in Section~\ref{bhStrength}).

It is also interesting to note that none of the black holes in these simulations reach the mass scales ($\sim10^{7}-10^{8}\,{\rm M_\odot}$) predicted from the observations (e.g. \citealt{magorrian1998}; \citealt{tremaine2002}). There are some possible causes for this discrepancy, several of which are discussed 
later in this paper, such as reduced accretion in the absence of mergers, low initial black hole mass, resolution effects and large-scale gas accretion. Investigating the 
final issue would require the inclusion of a hot gas halo (e.g. \citealt{sinha2009,moster2011,wurster2013}) and/or 
the simulation of large-scale accretion onto the galaxy (e.g. \citealt{moster2012}), and is beyond the scope of this paper.

\subsection{Simulations with supernova \& AGN feedback}
\label{DGsnplusagn}

We now investigate the effect of combining both supernova and AGN feedback in our simulations. 
Fig. \ref{fig:DGsfrsx} shows a mild reduction in the star formation rate in the simulations including both strong AGN 
and supernovae (SAGN\_SN), compared with the supernova only case (SN). This is due to AGN winds coupling to a 
supernova-fuelled hot gas halo reducing cooling and subsequent star formation (the effect is slightly more prominent in 
lower resolution simulations; not shown). This has been verified by examining the least dense particles in the 
simulations and it was found that gas driven out to large radii by strong AGN feedback is missing in simulations also 
including supernova feedback. The effect is not seen in the KAGN\_SN models as there the AGN drives a much weaker wind 
due to the larger mass being heated.

\begin{figure*}
\begin{center}
\includegraphics[width=14cm]{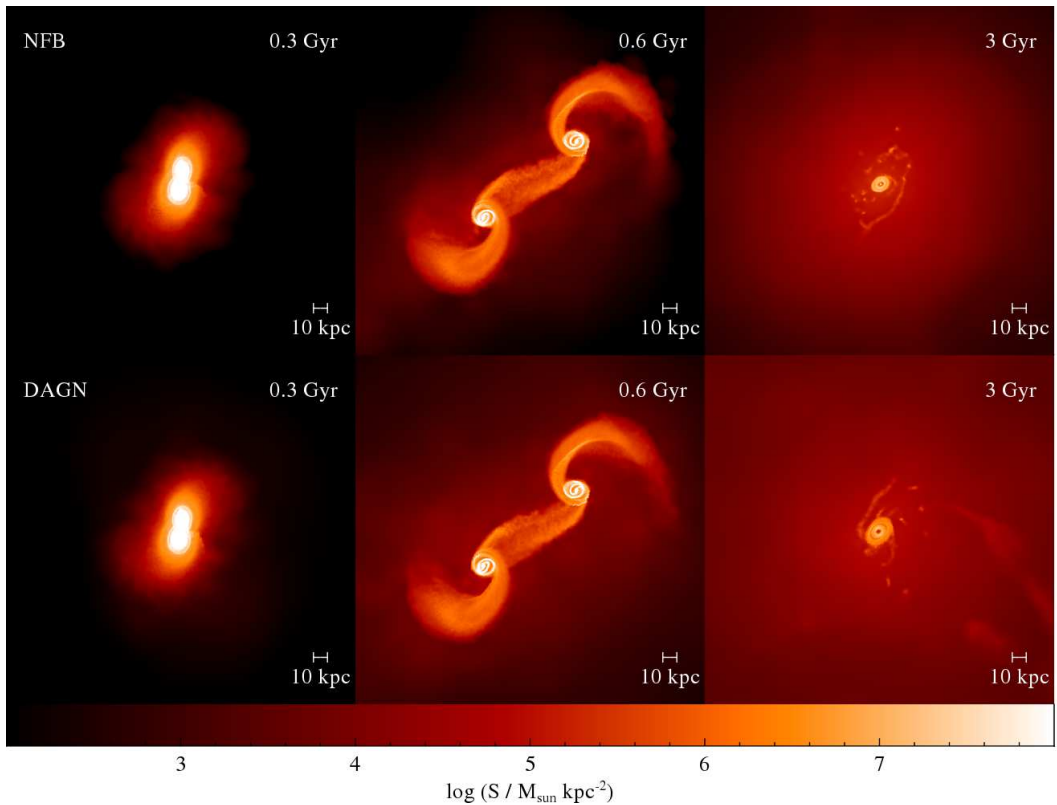}
\end{center}
\caption{Projected gas surface density maps for merging galaxies in the NFB (top panels) and DAGN simulations 
(bottom panels). Columns show snapshots taken at times, $t=\,0.3, 0.6$ and $3\,{\rm Gyr}$, viewed perpendicular to the merger plane. 
Panels are $250\,{\rm kpc}$ on a side and colours represent surface density on a log scale in units $\rm M_\odot\,kpc^{-2}$. Viewed at 
$t=\,0.3\,{\rm Gyr}$, the upper (lower) galaxy has entered from the right (left) with its rotation axis tilted (parallel) 
with respect to the line-of-sight and swings around to appear as the lower (upper) galaxy at $t=\,0.6\,{\rm Gyr}$ producing 
tidal tails and bridges, before finally coalescing.}
\label{fig:MGpretty}
\end{figure*}

Semi-analytic galaxy formation models commonly assume that the effects of supernova and AGN feedback affect galaxy properties 
independently of one another
(e.g. \citealt{bower2006,croton2006,guo2011}). This approach has recently been brought into question by the work of 
\cite{boothschaye2013}, who found that the two feedback processes suppressed the star formation rate less than if 
they were assumed independent. For a more quantitative analysis we calculated the feedback suppression `$\chi$' parameter 
as used in \cite{boothschaye2013}, where $\chi$ is defined as
\begin{equation}
\chi=\frac{S_{\rm AGN+SN}}{S_{\rm AGN}\times S_{\rm SN}},
\end{equation}
and $S_{\rm X}$ is the suppression in the star formation rate due to process ${\rm X}$.
The value of $\chi$ quantifies the level of interaction of AGN and supernova feedback which results in the suppression of 
star formation. A value of $\chi >1$ ($\chi <1$) shows that the two feedback processes have amplified (weakened) each others' 
ability to reduce the star formation rate and $\chi =1$ indicates that the two processes are independent. We find 
$\chi \approx 1$ at all times for KAGN runs and  $\chi \simeq 1$ rising to $\chi\simeq 1.15$  at late times for the SAGN and DAGN runs. 
This indicates that the interaction between the two feedback processes is of minimal importance, even for the strong feedback simulations 
where a coupling is observed. This weak 
amplification disagrees with the value of $\chi\sim 0.3$ found by \cite{boothschaye2013} for systems of 
$M\sim 10^{12}\,M_\odot$ such as this. However as this work considers one high resolution object in detail whereas 
\cite{boothschaye2013} perform a cosmological simulation, this finding is not necessarily a direct contradiction. Both the 
environment of the galaxy and how well the outflow is resolved are likely to be important in determining the strength of this effect.

The lower panel of Fig. \ref{fig:DGbhMassEvo} shows the ratio of BH masses for the main AGN models with and without 
supernovae over the course of the simulations. Regardless of model, the black hole mass is mildly enhanced by the 
action of supernovae after $1\, {\rm Gyr}$. We have verified that this effect is due to a combination of reduced star formation leaving more gas to accrete 
and winds from the star-forming disc feeding the central BH. 

\vspace{1em}

In summary, the reduction in star formation rate in our isolated galaxy is dominated by the effect of supernovae, 
whilst AGN have very little impact regardless of whether the feedback processes are simulated 
in isolation or in tandem. The final black hole mass is found to differ between models and consistently shows a mild 
increase as a result of the inclusion of supernovae.

\section[]{Merging disc galaxies}
\label{mergeGalaxies}

\begin{figure*}
\begin{center}
\includegraphics[width=17.5cm]{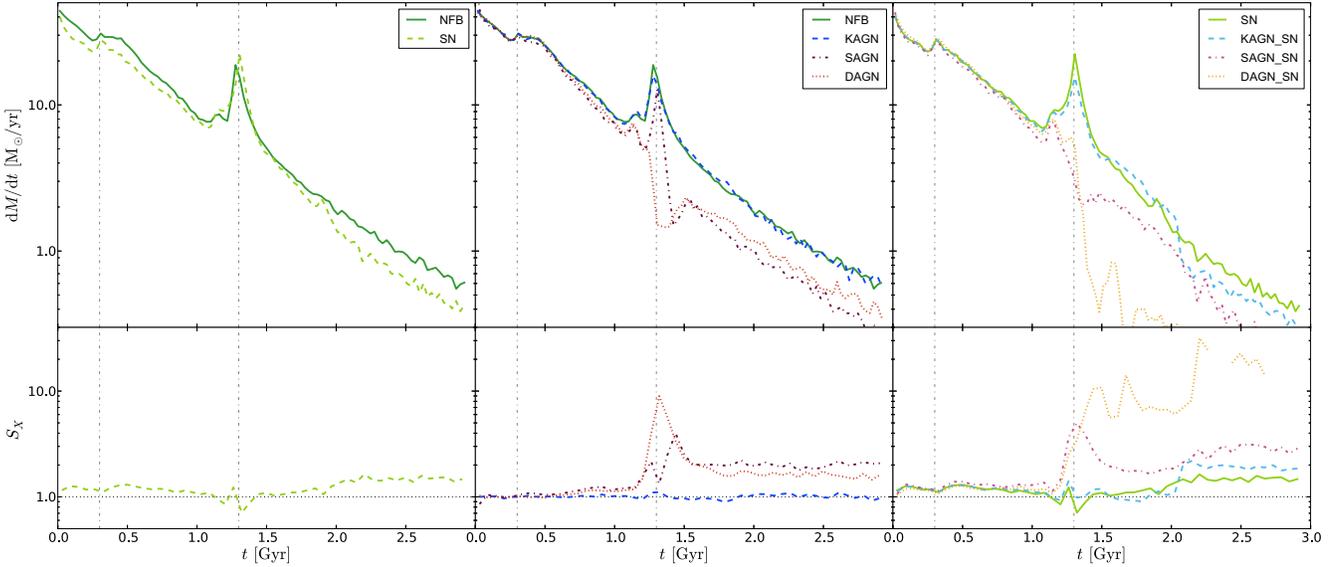}
\end{center}

\caption{Star formation rates and suppression factors for merging galaxy simulations with supernova feedback (left panels); 
AGN feedback (middle panels); and both (right panels). Vertical lines indicate the times of first and second passage. 
The suppression factor is the ratio of the feedback case to the no-feedback case. Strong AGN feedback has the largest 
impact on the star formation rate, eliminating the main starburst.}
\label{fig:MGsfrsx}
\end{figure*}

Having investigated the influence of supernova and AGN feedback processes on isolated systems, we now turn our attention to how 
they interact through a major merger. It is important to consider merging systems as they are thought to be a critical phase in 
galaxy evolution, driving both periods of high AGN activity and starbursts. It has been shown previously by \cite{SDH2005modelling} 
that the inclusion of AGN feedback significantly reduces star formation during the starburst. Here we re-investigate this result to 
establish whether this is a generic feature seen in our runs.

\subsection{The merger scenario}

We have chosen to simulate a 1:1 major merger between identical disc galaxies as described in Section \ref{ICsection}. This 
extreme choice was made primarily to allow for easier comparison with the isolated galaxy simulations and with previous work. 
We start the merger with an initial separation of $150\, {\rm kpc}$, and a perpendicular impact parameter of $25\, {\rm kpc}$. 
The galaxies initially have approximately zero total energy, corresponding to a net velocity for each galaxy of 
$185\, {\rm kms^{-1}}$ in the centre of mass frame, directed parallel to the separation axis. One of the galaxies is 
rotated by $30^{\circ}$ with respect to both the axis of symmetry and the plane of the merger. 
This merger geometry is intentionally similar to earlier work (e.g. \citealt{SDH2005modelling,debuhr2010}). Our choice of 
merger set-up results in the galaxy centres experiencing two passes, coalescing on the second, and merging at around 
$t\sim 1.5\,{\rm Gyr}$. Three key phases of the merger are illustrated in Fig. \ref{fig:MGpretty}, which shows the column 
density of gas for the first pass, largest subsequent separation and final state, for the NFB and DAGN simulations. 

At $t=3\,{\rm Gyr}$ the merging system has had time to relax back into an equilibrium state. 
The stellar population of the remnant is a slowly rotating oblate ellipsoid with sphericity, $s(=c/a)=0.49$; elongation, 
$e(=b/a)=0.86$; and triaxiality, $T[=(a^2-b^2)/(a^2-c^2)]=0.33$ where $a>b>c$ are the square roots of diagonal elements in the 
mass distribution tensor (see \citealt{bryan2013}). The abundance of gas at late times is model-dependent, but takes the form 
of a slowly rotating extended halo with a rotating ($v_{\rm c}\sim 200\,{\rm kms^{-1}}$) dense gas disc at its very centre. 
The central disc is formed at a small angle to the plane of the merger due to the initial tilt applied to one of the galaxies 
(Fig. \ref{fig:MGpretty}). All merger remnants have low stellar angular momentum and are classified as `slow rotators' under 
the scheme applied to the ATLAS 3D galaxies (e.g. \citealt{emsellem2007, bois2011}) independent of any feedback processes.

\subsection{Mergers with supernova feedback only}

\begin{figure}
\begin{center}
\includegraphics[width=\columnwidth]{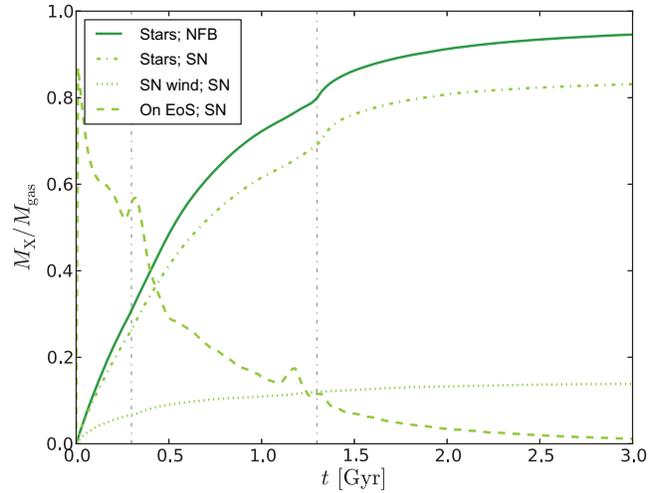}
\end{center}
\caption{Fractional mass in new stars and gas in various phases for simulations of high-resolution disc galaxy mergers with and 
without supernova feedback. For the NFB case only the stellar mass is shown, while for SN we additionally plot the supernova-heated 
particles and gas on the equation of state. All masses are shown as a fraction of the initial gas mass. As 
in the isolated case, supernovae reduce the amount of gas on the equation of state and stellar mass formed.}
\label{fig:MG_sn_Mphase}
\end{figure}

\begin{figure*}
\begin{center}
\includegraphics[width=17.5cm]{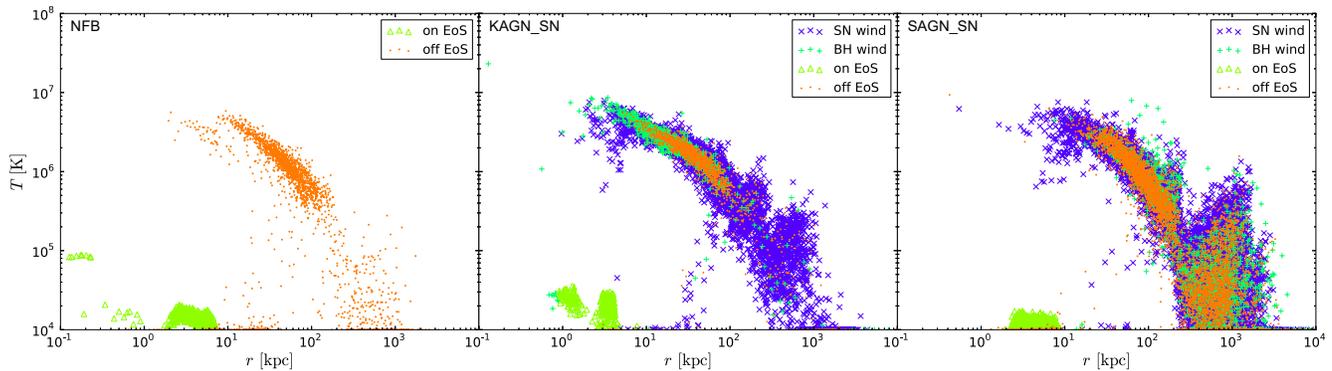}
\end{center}
\caption{Final temperature distributions versus radius for gas particles in the NFB, KAGN\_SN and SAGN\_SN (left to right) 
simulations, calculated at $t=3\, {\rm Gyr}$. Symbols indicate the `tags' applied to gas particles. The predominant trend in all 
panels is the same, indicating that the dynamical effects of the merger cause much of the heating and redistribution of the gas. 
However, the runs with strong AGN feedback (SAGN\_SN is shown here) produce a larger amount of hot gas ($T>10^5\, {\rm K}$) at 
large radii ($r>100 \, {\rm kpc}$).}
\label{fig:MGtprofSN}
\end{figure*}

We begin, as with our investigation into isolated galaxies, by evaluating the effect of the supernovae. 
The evolution of the star formation rates for the main simulations are shown in Fig. 
\ref{fig:MGsfrsx} as well as the suppression factor relative to the no-feedback case. For the NFB and SN models, an increase in 
the rate is seen for both passes ($t\sim 0.3$ and $1.3\, {\rm Gyr}$), however the burst at the time of second passage is 
substantially more significant, as found in previous studies when the galaxies contain a bulge (e.g. 
\citealt{mihos1996, SDH2005modelling, debuhr2010}). The primary burst takes place just as the galaxies coalesce, when the gas 
discs collide most strongly. As in the isolated case, we find that supernovae reduce the star formation rate throughout the 
majority of the simulation. However, star formation through the starburst is unchanged, suggesting that the supernova-heated 
gas is largely confined during this period.

Fig. \ref{fig:MG_sn_Mphase} shows the fraction of initial gas mass turned into stars versus time for the NFB and SN simulations. 
We also plot the fraction heated by supernovae and gas currently on the equation of state for the latter case. Almost all of the 
gas is converted into stars by the end of the simulation in the NFB run while supernovae reduce this fraction by $\sim 13$ per 
cent, similar to that seen in the isolated galaxy simulations. Supernovae cause an additional reduction of mass on the equation 
of state early on, reducing star formation and preserving gas. The amount of gas on the equation of state temporarily increases 
twice, coinciding with the timing of the galaxy passages that cause an increase in gas density during the collision. This leads 
to the enhanced star formation rate that acts to reduce the amount of gas on the equation of state once more. The amount of gas 
previously heated by supernovae shows only a weak change at the times of the merger passages as it has escaped to areas of 
lower density. The number of particles tagged as heated is, as in the isolated simulation, only a small fraction of the number
 of stars formed because the majority quickly rejoins the equation of state without escaping the disc.

We present in Fig. \ref{fig:MGtprofSN}, the final temperature distributions of gas particles for the NFB, KAGN\_SN and SAGN\_SN 
simulations (left to right). Here we discuss the effect of supernovae using the KAGN\_SN particle temperature distribution as a 
proxy for SN, to avoid duplication as they are remarkably alike. The distributions for NFB and SN are very similar, producing a 
halo of diffuse, hot gas that extends to $\sim 1\, {\rm Mpc}$ from the halo centre. This indicates that the dominant cause 
of heating is gravitational in origin and a consequence of the merger. The importance of dynamical effects in the 
formation of the hot halo is corroborated by the increase in gas fraction in the halo. This is $53$ per cent at the end of the 
simulation when feedback is omitted compared to $0.02$ per cent for the isolated galaxy 
(fractions with supernovae increase to $90$ and $48$ per cent respectively). 

The final gas density profiles are shown in Fig. \ref{fig:MGprofSN}. Profiles are shown at $t=3\,{\rm Gyr}$ and centred on the 
centre of stellar mass; also shown are the ratios of the density profiles compared to the NFB simulation. 
The SN profile shows only a small increase in gas density beyond $10\, {\rm kpc}$ due to a reduction in star formation.

\subsection{Mergers with AGN feedback only}
\label{mergerAGNonly}

\begin{figure*}
\begin{center}
\includegraphics[width=17.5cm]{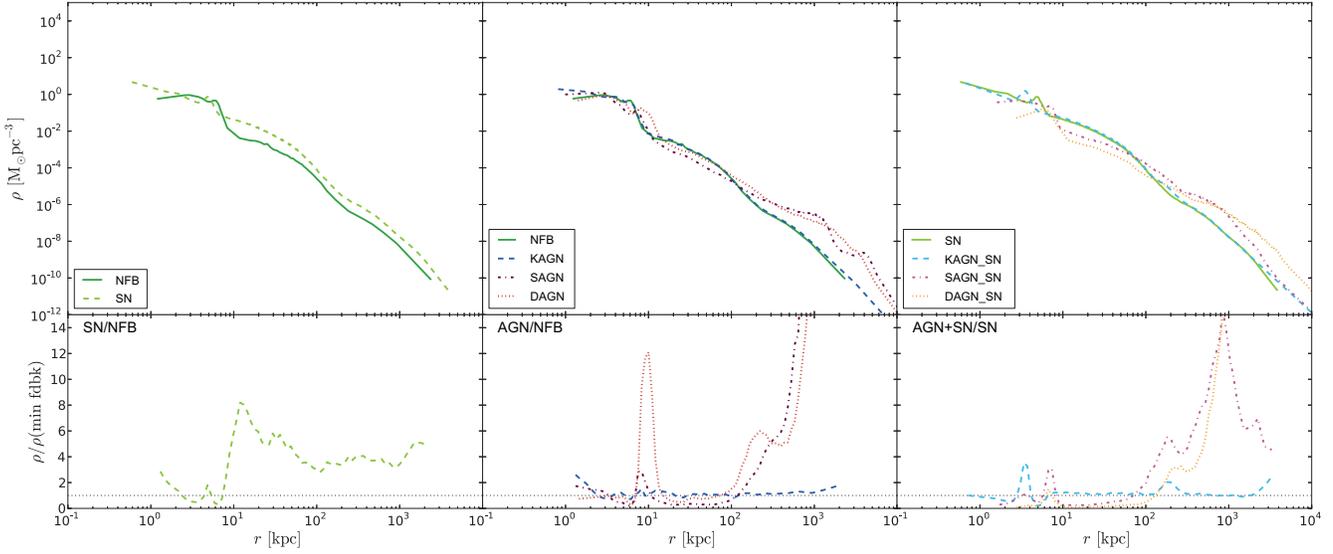}
\end{center}
\caption{Final gas density profiles for the main set of simulations, grouped into runs with no AGN (left panels), AGN only 
(centre panels) and supernovae plus AGN (right panels). The lower sub-panels show the ratio for each model relative to the 
NFB run in the left and centre panels, and the SN run in the right panel. Both supernovae and strong AGN feedback affect the 
gas distribution, with the latter pushing gas out to larger radii.}
\label{fig:MGprofSN}
\end{figure*}

In stark contrast to the isolated galaxy case we find that in merging systems, the AGN can have the dominant effect on the 
star formation rate when the feedback is strong (SAGN and DAGN; Fig. \ref{fig:MGsfrsx}). While the KAGN model has no impact on the star formation rate, the SAGN and DAGN models significantly reduce the amount of stars formed during and after the merger. In the DAGN case, the AGN also eliminates the starburst; if generic (and not just for our particular merger configuration) it conradicts observational evidence linking mergers with starburst activity (see e.g. \citealt{alexander2012} and references therein). Over the whole simulation, the weaker kernel feedback (KAGN) reduces the final mass in stars by only $5$ per cent, 
by reducing the starburst duration and strong feedback by $\sim25$ per cent. We investigate this further in Section \ref{bhStrength}.

The wide disparity in the power of the AGN to influence its environment is likely due to the differing temperatures which 
gas is heated to. At $10^8\, {\rm K}$ (the strong feedback heating temperature) radiative cooling has less of an effect, 
as the cooling time increases as $\sqrt{T}$ when the emission is dominated by thermal Brehmsstrahlung. Therefore, by spreading 
the available feedback energy over many particles neighbouring the black hole and depositing any available energy promptly, 
the kernel-weighted method heats more gas to lower temperatures where it can cool more easily. The centre and right panels of 
Fig. \ref{fig:MGtprofSN} show that whilst kernel feedback removes gas from the equation of state and up to temperatures of 
$\sim 10^5\, {\rm K}$, strong feedback heats much more gas to higher temperatures and larger radii (up to $10^6\, {\rm K}$ at 
$10^3\,{\rm kpc}$), resulting in less gas on the equation of state.

Final gas density profiles (middle panels of Fig. \ref{fig:MGprofSN}) show a redistribution of gas out to large radii in the case of 
the strong feedback models. The kernel-weighted feedback method makes very little difference to the gas distribution, 
causing only a slight increase in gas at $r\lesssim 10\,{\rm kpc}$. The strong feedback methods however redistribute gas out to much 
larger radii, $r> 10^2\,{\rm kpc}$, with an order of magnitude larger density than stars at that radius, 
increasing the mass of gas at $r>10^3\, {\rm kpc}$ by two orders of magnitude from $2\times 10^7$ to $2.5\times 10^9\,{\rm M_\odot}$. 
This does not, however, significantly change the baryon fraction within $R_{200}$ compared to the NFB case. AGN have little effect 
on stellar density profiles, as found for isolated galaxies, due to the subdominant mass in new stars and the large amount of 
stellar mixing which occurs though a merger. Furthermore, gas removed to lower densities by strong AGN feedback is less likely 
to return and will never form stars over the course of the simulation.

Evolution of the black holes with time for the three main AGN models are shown in Fig. \ref{fig:MGbhMassEvo}. The top panels show the average mass accretion rates (in units of the Eddington rate), while the middle panels show the integrated masses of the two black holes as well as their sum. The black holes undergo a spike of growth at the 
times of the first and second passes, forming a tight binary system during the latter which has not coalesced by the end of the 
simulation (black holes are found to merge more readily in low resolution simulations). Correspondingly there is a large 
increase in the net amount of feedback energy deposited compared to the isolated case, which in itself suggests the AGN will 
influence its environment more strongly. The AGN activity peaks are coincidental with merger passes, as the latter drives an 
inflow of gas, increasing the density around the black hole, and thus boosting the accretion rate.

\subsection{Mergers with supernova \& AGN feedback}
\label{mergersSNAGN}

\begin{figure*}
\begin{center}
\includegraphics[width=17.5cm]{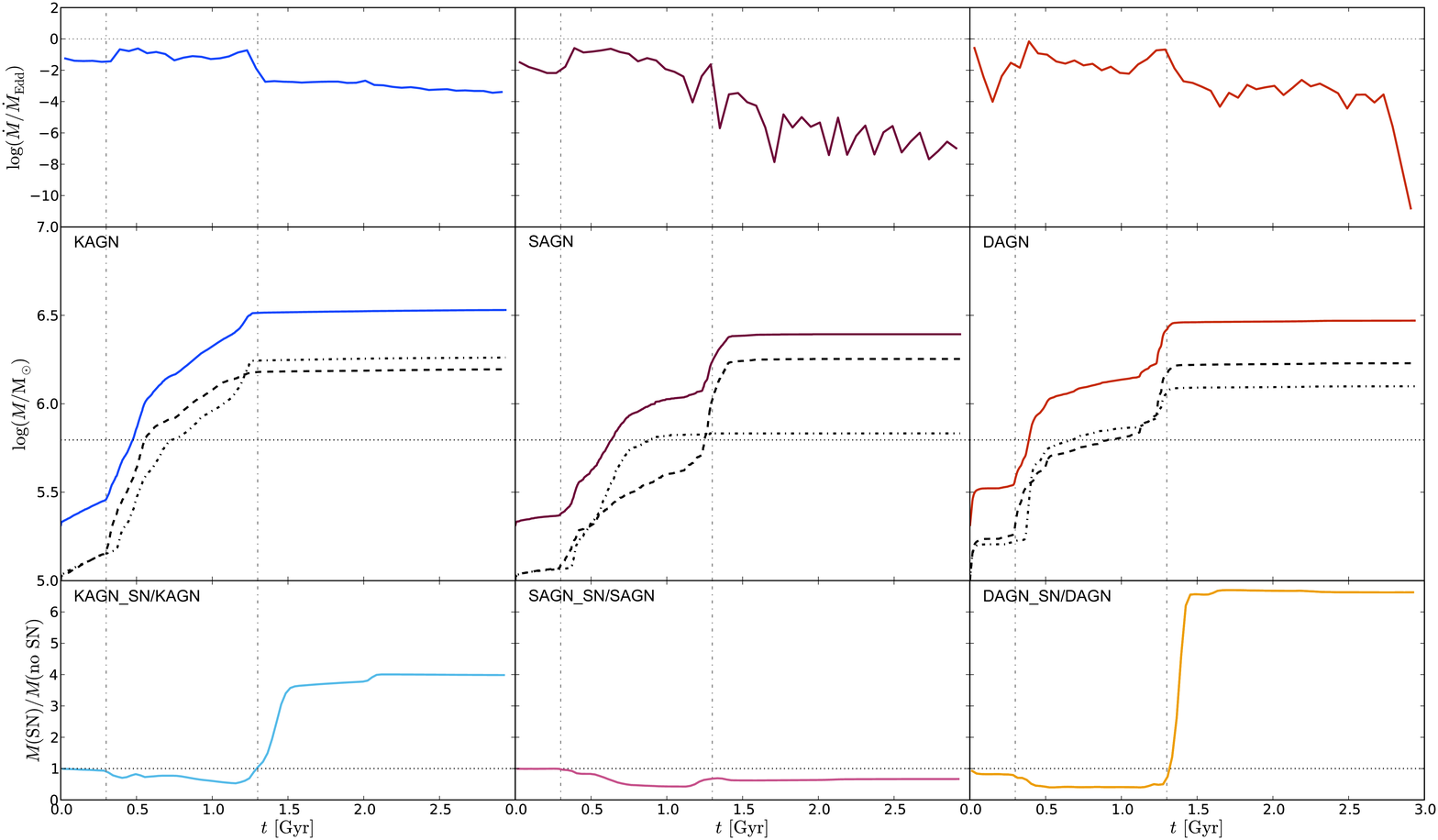}
\end{center}
\caption{Evolution of the black holes with time through a merger for the fiducial AGN models; left to right, KAGN, SAGN and DAGN respectively. 
The top panels show the binned total mass accretion rate in units of the Eddington rate, onto both of the black holes, for models without supernovae. In the middle panels, the integrated `internal' black hole masses are shown (dotted and dashed lines), with the solid line representing the sum of the two. Lower panels show the total 
black hole mass for each AGN model with supernovae relative to the no-supernovae equivalent. The horizontal black dotted 
line in the central panel shows the threshold mass above which black hole recentring is halted, whilst elsewhere it denotes a value of unity. Boosted growth rates can be seen 
at the times of first passage and coalescence of the merger. The inclusion of supernovae increases the final black hole mass 
when the two black holes do not merge (as is the case with KAGN and DAGN).}
\label{fig:MGbhMassEvo}
\end{figure*}

The previous sections have shown that AGN can play a more significant role than supernovae in determining the 
star formation rate during a merger. We can see from the star formation rates and ratios in Fig. \ref{fig:MGsfrsx} that strong
AGN feedback (SAGN and DAGN models) is even more effective at reducing star formation for the majority of the simulation with the starburst now being terminated in both cases. The SN and AGN both provide additional thermal energy to hot halo gas, increasing its cooling time and thus reducing the amount of material joining the equation of state.

For a more quantifiable measure of the combined suppression effects of the feedback processes, we again calculate the $\chi$ 
parameter. We find, as in the isolated galaxy simulations, $\chi$ values consistent with unity prior to coalescence but 
rising to $\sim 3\, (1.4)$ at late times for the strong (kernel-weighted) models respectively. This, as discussed in the context of 
isolated galaxies, is a departure from the findings of \cite{boothschaye2013}. However, we note that $\chi$ values 
calculated after the merger are prone to noise due to the low-level of star formation at this time.

The coupling of the AGN outflow to the supernova-fueled hot gas halo, observed in both our isolated and merging galaxy 
simulations, raises the question of whether such a halo ought to have been included in our initial conditions. Such an 
approach has been adopted by recent galaxy merger studies (e.g. \citealt{sinha2009,moster2011}) and may have affected our 
findings through the influence of feedback on late-time gas accretion from the halo. However, since we compare our galaxy 
models primarily to each other rather than observations any effect should be reduced, and simulations including supernovae 
quickly generate a halo in any case. Furthermore, omitting this component has allowed a high resolution study on the role 
of feedback processes which is more easily compared with preceding studies.

Supernovae have a stronger effect on black hole mass in mergers compared to the isolated galaxy simulations. In the KAGN\_SN 
and DAGN\_SN simulations the reduction in star formation resulting from the inclusion of supernovae results in more gas in 
the galaxy after coalescence to feed the black hole (Figs. \ref{fig:MGprofSN} and \ref{fig:MGbhMassEvo}). The heating from 
supernovae therefore acts to increase growth by reducing star formation and leaving more surrounding gas at late times to be 
accreted by the black hole. This trend is not seen in the SAGN\_SN simulation because the black holes merge promptly at 
coalescence, resulting in a comparatively reduced total accretion rate as the merged black hole stablilises at the centre of 
the system.

\vspace{1em}

In this section we have shown that, in contrast to simulations of isolated galaxies, it is possible for AGN to play the 
dominant role in suppressing star formation in a merger scenario. The method by which feedback energy is deposited is 
crucial in determining the strength of suppression; whilst models incorporating strong feedback eliminate the starburst, 
the weaker kernel-weighted mode merely reduces its duration. Supernovae still play a role in suppressing the star formation 
during the majority of the simulation, but are found to be incapable of reducing the strength of the starburst by themselves.

\section[]{A closer look at the models}
\label{discussion}

Our main simulations have allowed us to analyse the role feedback processes play in disc galaxies for a range of existing models. 
We now go on to look into numerical effects and the impact of our chosen parameters. As the high resolution achieved in 
these simulations is currently out of reach of typical cosmological simulations, we begin by investigating the impact of 
employing a lower mass resolution in Section \ref{res_effects}. The effect of increasing the galaxy gas fraction to better 
represent the high redshift systems in which most quasar growth and mergers are believed to occur is examined in Section 
\ref{gasFracRuns}. To analyse the consequence of choosing an initial black hole mass of $10^5\,{\rm M_\odot}$, we perform 
tests in which we set this to a larger value in Section \ref{bhMassRuns}. 

We also discuss issues raised whilst performing this work and further investigate aspects of the feedback models. 
Simulations of isolated galaxies showed that black holes appear to self regulate, slowing their growth as found by \cite{DMS2005}. 
However this may equally be caused by star formation consuming the gas which would otherwise feed the black holes; we 
study this further by artificially turning off star formation in Section \ref{noSFruns}. A further investigation provoked by the 
main body of this work originated in the finding that the temperature to which we heat gas in AGN feedback is critical. We extend 
our investigation to varying supernova temperature in Section \ref{snTemp}. Finally, we disentangle the accretion and feedback 
elements of the \cite{DMS2005} and \cite{boothschaye2009} models incorporated in the main investigation to confirm that the 
AGN feedback temperature is the main feature driving the change in the star formation rate in Section \ref{bhStrength}. 

\subsection{Numerical resolution}
\label{res_effects}

In order to evaluate the effect of changing the simulation mass resolution, we re-run the main suite of simulations with one tenth 
of the number of particles but with all other galaxy parameters held constant (see Table \ref{tab:Resparams}, comparing Low versus High resolution). We begin by 
probing the differences in the star formation rate for simulations run at lower resolution.

\begin{figure}
\begin{center}
\includegraphics[width=\columnwidth]{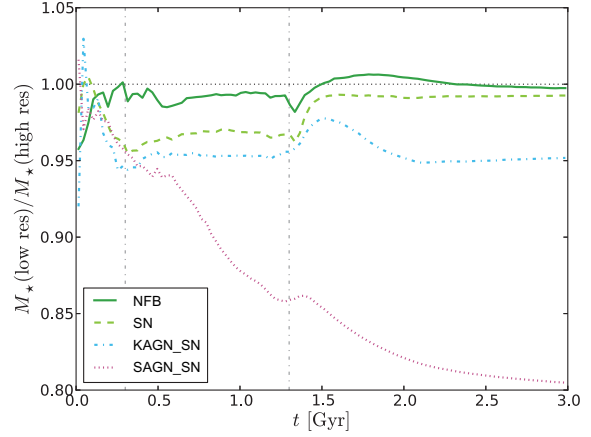}
\end{center}
\caption{Ratio of stellar mass formed during simulations performed at high and low resolution for NFB, SN, KAGN\_SN 
and SAGN\_SN models in merging galaxies. The SAGN\_SN simulation exhibits the strongest resolution 
dependency due to the difficulty in resolving a columnated energetic outflow at low resolution. Under-resolving 
the outflows increases the amount of gas with which they interact with and heat.}
\label{fig:Mstarresratio}
\end{figure}

The star formation rates in simulations of isolated galaxies show a small decrease in star formation due to a reduced ability 
to resolve high-density clumps where stars form. However the dependence on resolution is slight, with around a $\sim 10$ per 
cent reduction, so we do not discuss this further. The resolution dependence of simulated mergers is more complex. 
Fig. \ref{fig:Mstarresratio} shows the ratio (low resolution/high resolution) of stellar mass formed against time for the 
NFB, SN, KAGN\_SN and SAGN\_SN simulations (simulations without supernovae show identical trends). 
As found in isolated galaxies, low-resolution simulations generally under-produce stars, leaving more gas. However, they 
show an {\it increase} in star formation during starbursts as this gas is compressed during the merger. 
The NFB, KAGN, SAGN and DAGN simulations produce $1, 5, 15$ and $17$ per cent 
less stellar mass over the course of the simulations respectively\footnote{We note that changing the simulation mass resolution 
does not directly modify the time interval or mass loading of feedback events for a given accretion rate in the strong feedback 
method because we decrease $N_{\rm heat}$ by the same factor by which we decrease the resolution (to $N_{\rm heat}=1$).}. The 
larger reduction in stellar mass for the strong feedback AGN models (DAGN and SAGN are very similar at lower resolution) is 
due to the powerful outflows interacting with more gas, as a result of the larger smoothing lengths in low resolution simulations.

\begin{figure}
\begin{center}
\subfigure{\includegraphics[width=\columnwidth]{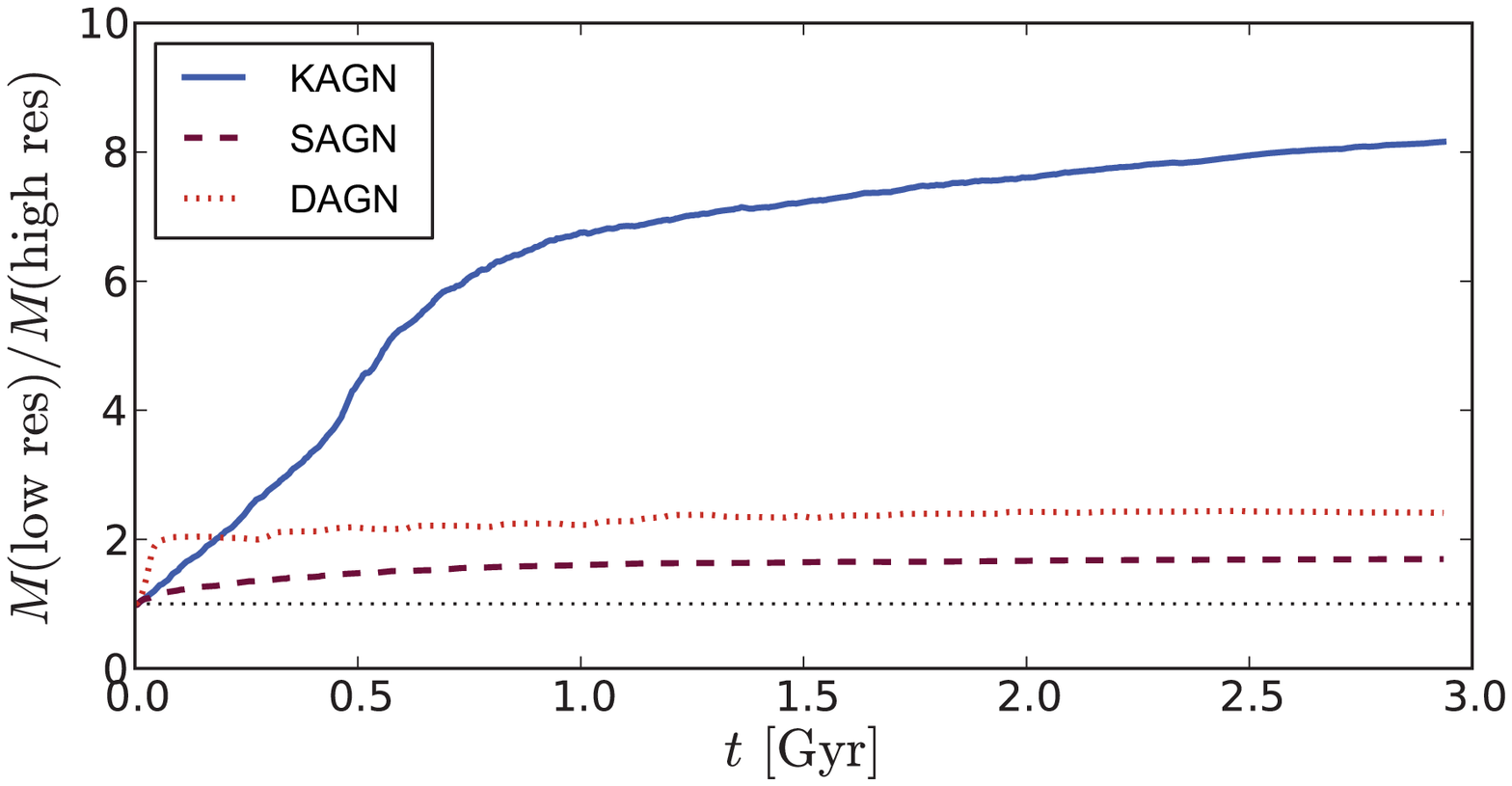}}
\subfigure{\includegraphics[width=\columnwidth]{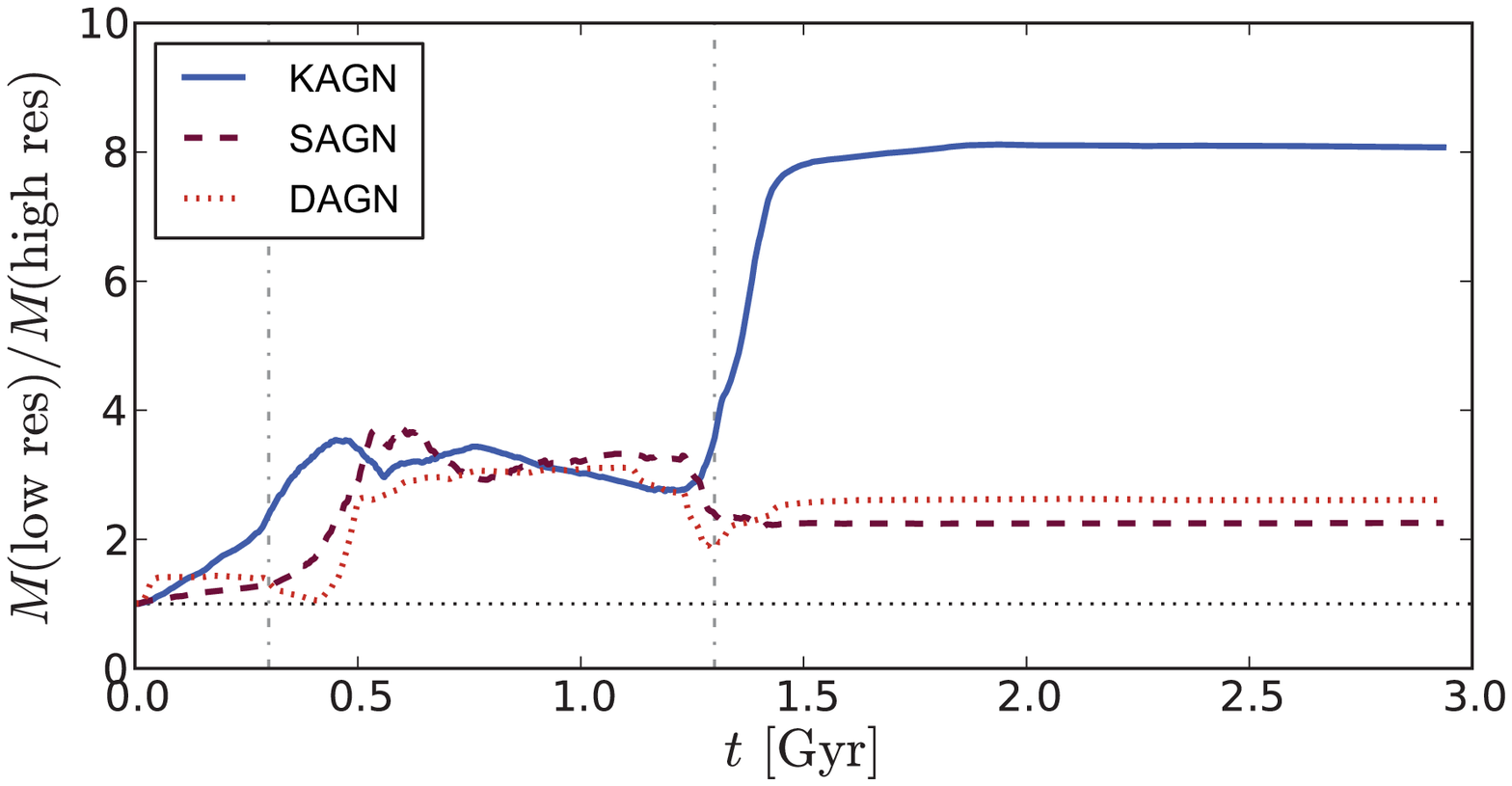}}
\end{center}
\caption{Total black hole mass ratio in low resolution over high resolution simulations in isolated (top) and merging (bottom) 
galaxies. For all models, black holes grow significantly more in low resolution simulations.}
\label{fig:BHmassRes}
\end{figure}

The black hole mass in low resolution simulations, plotted as a fraction of the high resolution mass is shown in Fig. 
\ref{fig:BHmassRes} for isolated galaxies (top panel) and mergers (bottom panel), for the main AGN simulations. All AGN models 
produce a larger black hole mass in low resolution simulations. This resolution dependence is largely due to more poorly resolved 
accretion, meaning that in low resolution simulations a black hole can accrete much more mass before capturing a particle and 
lowering the local density. The effect is even more severe in the KAGN model, where the typical heating temperature 
decreases in the low resolution case as the kernel mass increases.

For the low and high resolution simulation sets we chose to vary the particle number whilst holding the gravitational softening length constant. As the softening length is linked with the minimum smoothing length ($h_{\rm min}=2.8\, \varepsilon_{\rm soft}$) and DAGN accretion radius ($r_{\rm acc}=\varepsilon_{\rm soft}$) this fixes the typical physical scale for the AGN methods regardless of resolution (we note both methods therefore have an undesirable dependence on spatial resolution). It is, however, more common in cosmological simulations to vary the softening length to avoid numerical effects such as two-body relaxation (\citealt{power2003}). Additionally it is unclear whether the resolution effects on the star formation and black hole mass are a generic trend. 

In order to address these issues we perform two additional simulation sets (see Table \ref{tab:Resparams}) with particle number equal to the Low resolution and a decade lower (Very-Low; $N=7\times 10^4$, $N_{\rm heat}=0.1$ where applicable) respectively. The additional simulations are performed with softening lengths scaled with respect to the High resolution runs as $\varepsilon_2=(N_1/N_2)^{1/3}\varepsilon_1$; yielding softening lengths of $0.05$, $0.11$, $0.23\, {\rm kpc}$ for the High, Low and Very-Low runs respectively.

\begin{figure}
\begin{center}
\subfigure{\includegraphics[width=\columnwidth]{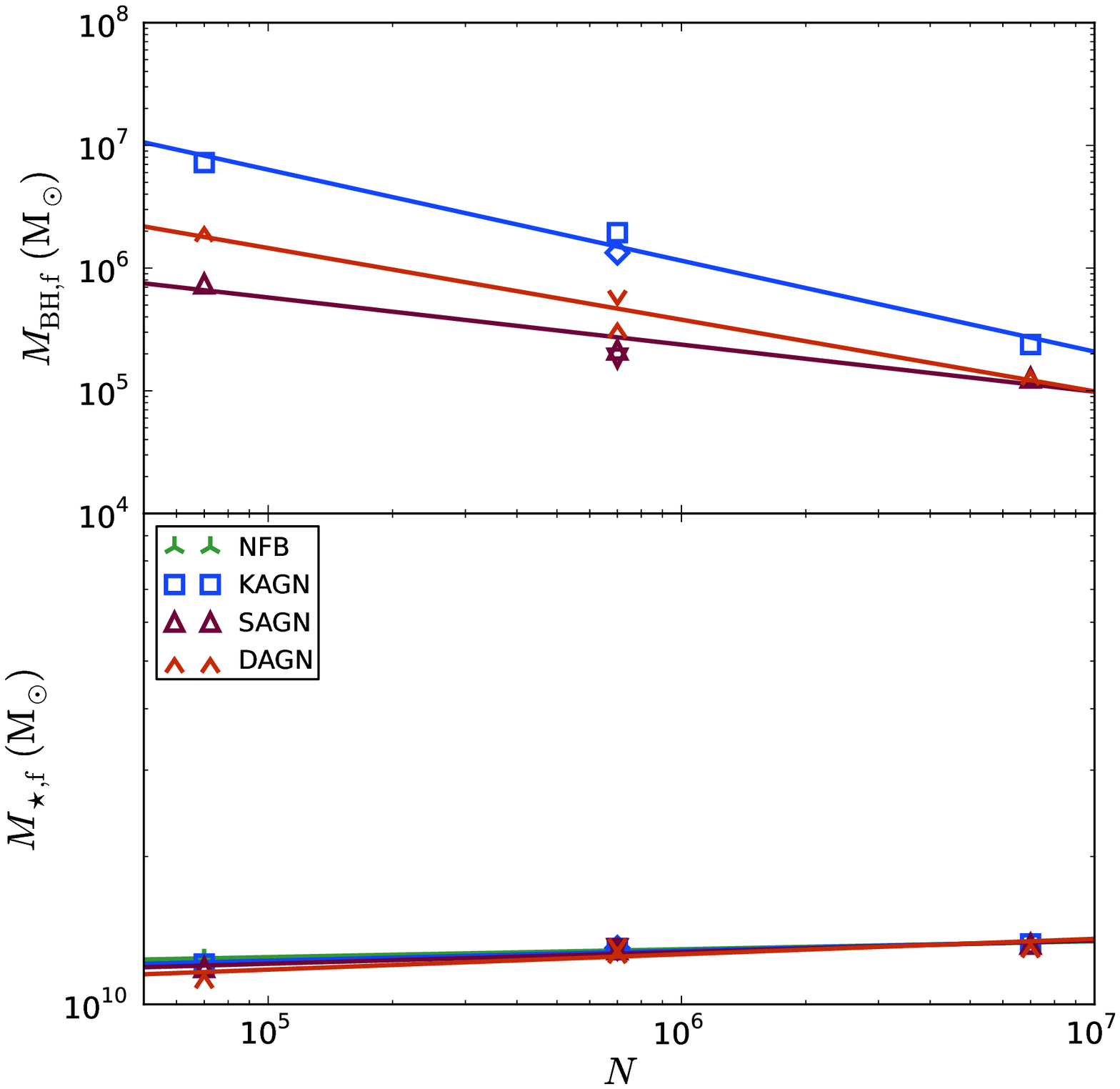}}
\subfigure{\includegraphics[width=\columnwidth]{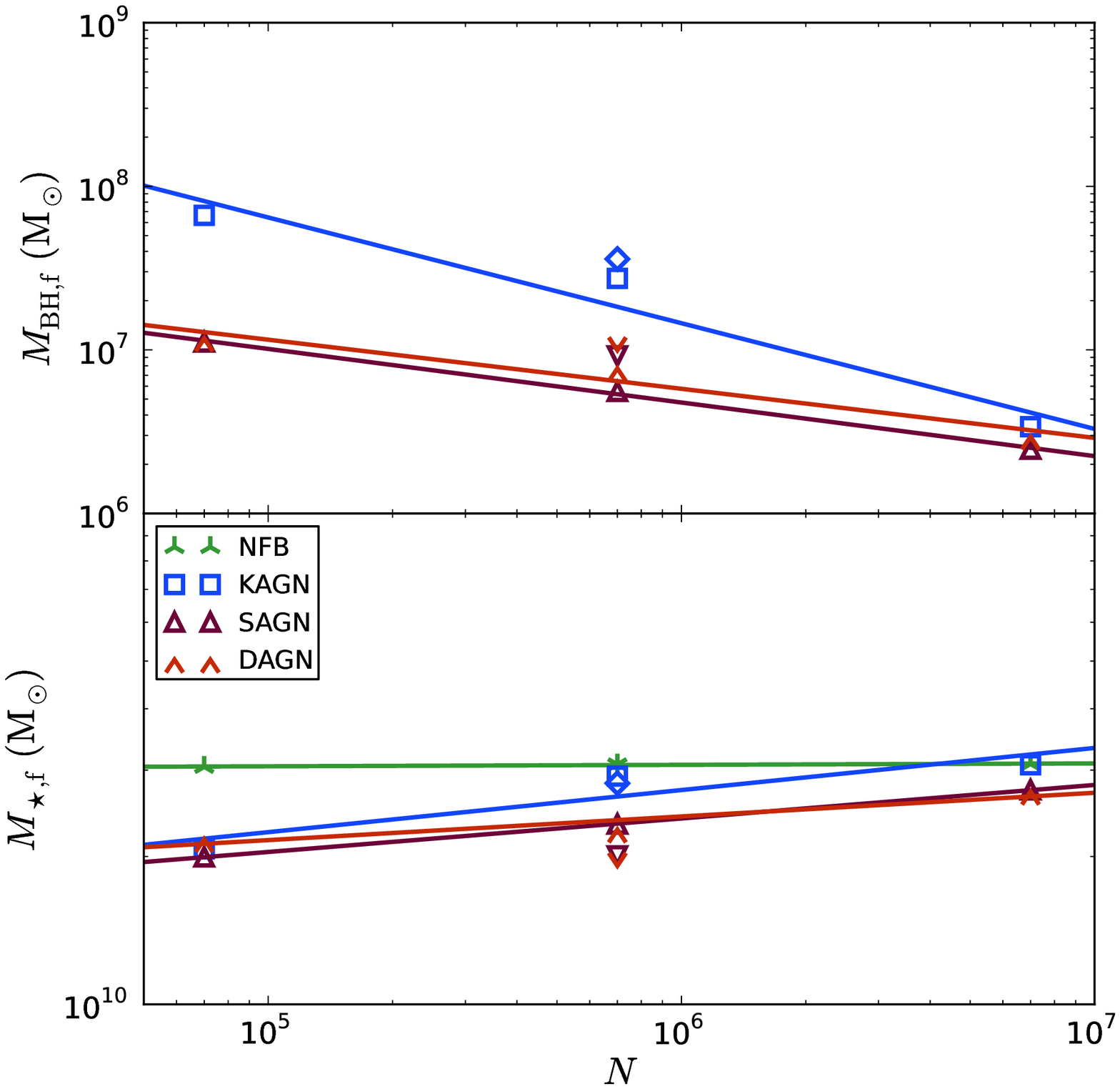}}
\end{center}
\caption{Total mass of black holes and stars formed (upper and lower subpanels respectively) against total number of particles at the end of isolated (top) and merging (bottom) galaxy simulations for the fiducial AGN models. Variant symbols denote the standard Low resolution simulations whereas the others have scaled softening lengths; lines are fits to the scaled-softening simulations.
A Very-Low resolution simulation confirms the trends of decreasing black hole mass and increasing stellar mass with higher resolution AGN simulations.}
\label{fig:ResScale}
\end{figure}

We present in Fig. \ref{fig:ResScale} the total mass of black holes and stars formed against total particle number for the extended simulation set performed with the fiducial AGN models (omitting supernova feedback). We find that performing an additional simulation at Very-Low resolution confirms the trends previously seen. Across models, lowering the resolution means more poorly resolved accretion, cold gas-clumps and AGN outflows which results in lower star formation rates and increased black hole masses as previously discussed. 
Increasing the softening length (in the $N=7\times 10^5$ simulations) makes negligible difference in the NFB simulation where the reduction in stellar mass formed is very slight, however simulations including feedback generally exhibit a slightly reduced final black hole mass. This is due to the increased minimum smoothing length resulting in larger outflows which lower the accretion rate more effectively. The lower accreted mass leads to less feedback energy deposited into the gas in total and allows for a minor increase in star formation.
It is interesting to note that the black hole masses obtained using different methods appear to be converging, however verifying this would require even higher resolution simulations, which we leave for further study.
We also note that varying the number of SPH neighbours (e.g. to keep the weighted mass within the kernel constant) between resolutions may give a better scaling for the KAGN model, however such a change would also modify the hydrodynamical behaviour of the gas.

A more direct comparison with the black hole masses found in the work of \cite{SDH2005modelling} and \cite{DMS2005} is 
made possible using our Very-Low resolution simulations, as their runs used $8\times 10^4$ particles. In isolated galaxy 
simulations we find a black hole mass of $7.2\times 10^6\,{\rm M_\odot}$ which is broadly consistent with Springel et al. 
($\sim2\times 10^6\,{\rm M_\odot}$). However it is not surprising to find a discrepancy as our differing resolution, star formation routine, ISM model and 
disc gas fraction will all affect the black hole mass. In merging systems we again find a final black hole mass higher ($6.7\times 10^7\,{\rm M_\odot}$) 
than the value obtained in previous studies ($\sim 1-3\times 10^7\,{\rm M_\odot}$; \citealt{SDH2005modelling,johansson2009}). 
The simulations with an irregular merger geometry such as our own are at the lower end of this range, therefore the deviation is slightly larger in merging systems than isolated.

\subsection{Initial gas fraction}
\label{gasFracRuns}

\begin{figure*}
\begin{center}
\includegraphics[width=16cm]{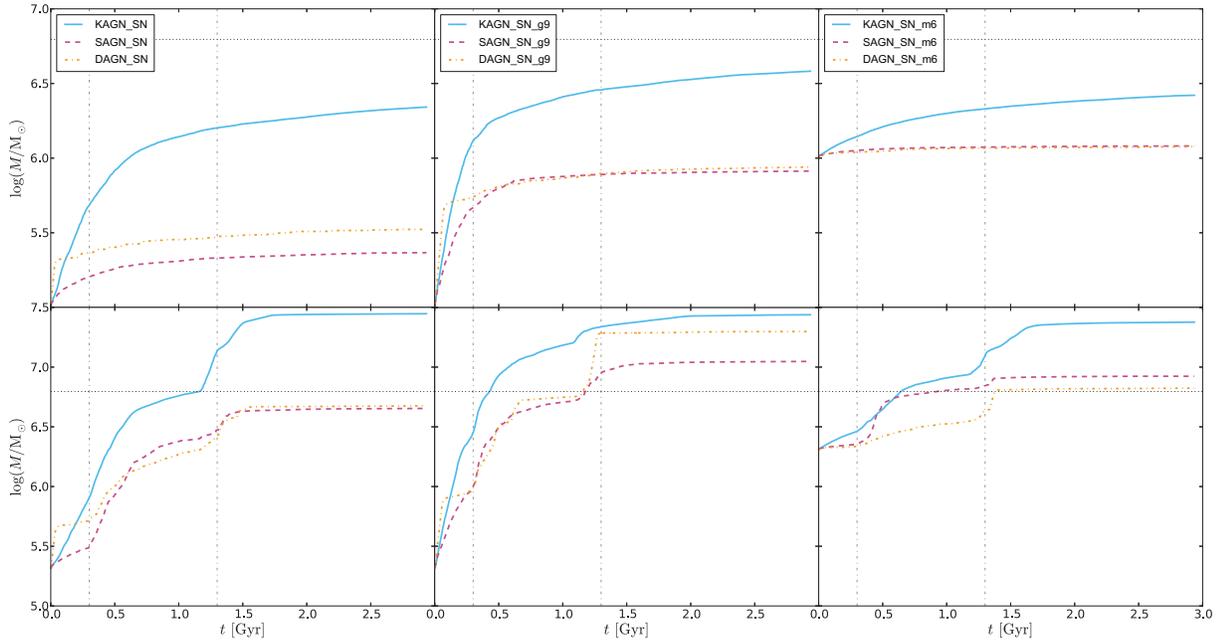}
\end{center}
\caption{Black hole mass throughout low resolution, isolated (top) and merging (bottom) galaxy simulations including 
the three primary AGN models. Columns (left to right) show: standard initial black hole mass and disc gas fraction 
($M_{\rm BH, i}=10^5\,{\rm M_\odot}$, $f_{\rm g}=0.3$); increased disc gas fraction ($M_{\rm BH, i}=10^5\,{\rm M_\odot}$, 
$f_{\rm g}=0.9$); and increased initial black hole mass ($M_{\rm BH, i}=10^6\,{\rm M_\odot}$, $f_{\rm g}=0.3$). The final 
black hole mass is relatively unaffected (to within a factor of 2-3) by changing either the disc gas fraction or initial mass. }
\label{fig:resBhMassGrid}
\end{figure*}

\begin{figure}
\begin{center}
\includegraphics[width=\columnwidth]{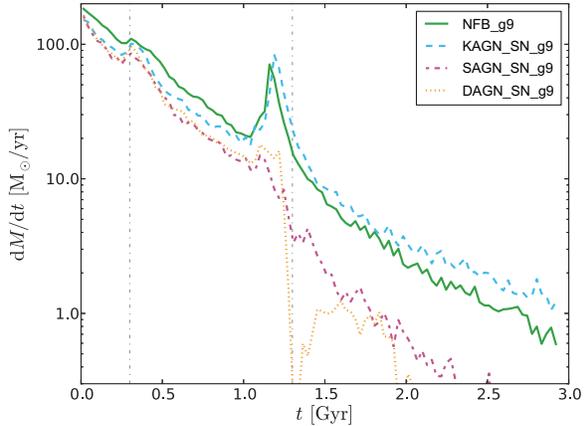}
\end{center}
\caption{Star formation rate throughout simulations of low resolution merging galaxies with an increased disc gas 
fraction $f_{\rm g}=0.9$ including the three primary AGN models and supernovae (KAGN\_SN\_g9, SAGN\_SN\_g9, DAGN\_SN\_g9). 
The shape and trends are the same as seen with the fiducial gas fraction, however the absolute rate is almost an order 
of magnitude larger.}
\label{fig:SFR_MG_g9}
\end{figure}

We now investigate how the supernova and AGN feedback processes interact in gas-rich discs with a gas fraction 
$f_{\rm g}=0.9$, more appropriate for a high redshift system. (The simulations are performed at low resolution 
for the standard AGN models to minimise CPU time consumption.) Fig. \ref{fig:resBhMassGrid} shows the black hole mass 
evolution for the low resolution fiducial models (left panels) and increased gas fraction (centre panels). 
This change has the effect of accelerating the coalescence of the galaxies/black holes. However, we find that the black 
holes in gas-rich systems cease growth at broadly similar masses to the fiducial simulations for all 
AGN models. This is perhaps surprising as we have shown that the black hole growth is very sensitive to the amount of gas 
available in its surroundings (Section \ref{mergersSNAGN}). This is explained by the increased level of star formation 
observed in the gas-rich galaxies (Fig. \ref{fig:SFR_MG_g9}) consuming the gas on the same timescale as in the fiducial 
galaxy and terminating black hole growth. Aside from the higher star formation rates seen across all times, the feedback 
models show the same trends in the star formation rate as in the case of the models with lower gas fraction. 
Supernovae dominate the star formation suppression in the isolated galaxies and through mergers the AGN model with weak 
feedback suppresses the star formation slightly, but it is shut off completely by the strong feedback models.

\begin{figure}
\begin{center}
\includegraphics[width=\columnwidth]{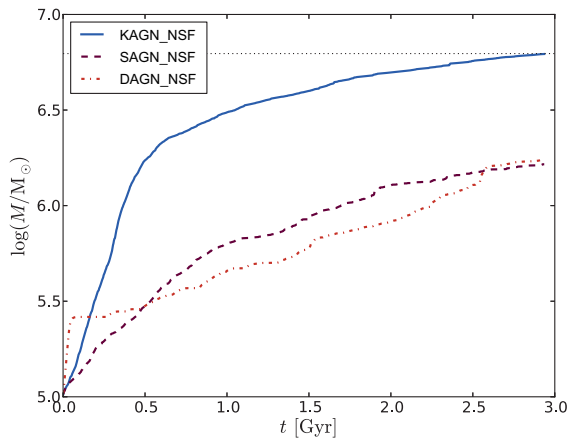}
\end{center}
\caption{Black hole mass evolution for isolated galaxies simulated with the fiducial AGN models and no star formation. 
This shows that the black holes have not self-regulated their own growth through feedback within $3\, {\rm Gyr}$. }
\label{fig:DGbhMassNOsf}
\end{figure}

\subsection{Initial black hole mass}
\label{bhMassRuns}

\begin{figure*}
\begin{center}
\includegraphics[width=17cm]{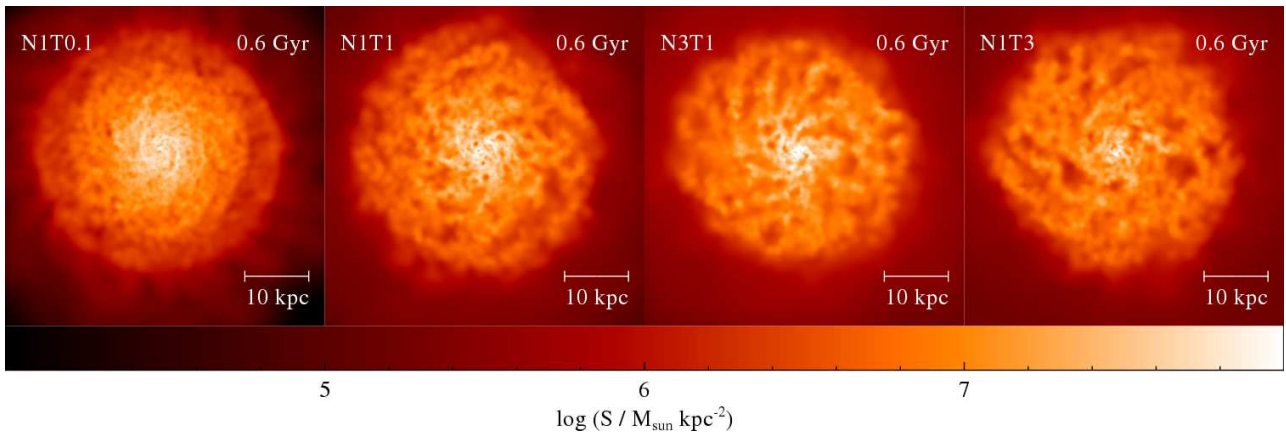}
\end{center}
\caption{Projected logarithmic face-on gas surface density maps for isolated galaxies simulated with different 
supernova feedback parameters at $t=\,0.6\,{\rm Gyr}$ (all simulations include Bondi-type kernel-weighted AGN 
feedback). Models shown have increasing (left to right) feedback strength: (KAGN\_SN\_)N10T0.1, N1T1, N3T1, N1T3. 
A larger heating temperature leads to an increased porosity of the gas disc.}
\label{fig:DGprettySN}
\end{figure*}

The initial black hole mass used in our main simulation runs is motivated by the approximate mass scale of 
primordial `seed' black holes formed through the direct collapse of zero metallicity gas (see e.g. \citealt{bellovary2011} 
and references therein) and the desire for compatibility with previous studies. This value is poorly constrained and 
well below the mass of a black hole predicted by observational relations for this galaxy, $M_{\rm BH}\sim 10^{7}\,M_{\odot}$ 
(\citealt{magorrian1998,tremaine2002,haringrix2004}). Here we investigate the sensitivity of the black hole mass 
evolution to the initial mass chosen. We perform additional low resolution simulations for the KAGN, SAGN and 
DAGN models with an initial mass of $M_{\rm BH, i}=10^{6}\,{\rm M_\odot}$. While this mass 
is substantially larger (10 times the fiducial mass) it is still lower than the predicted 
value and therefore still requires significant growth to reach this relation.

We find that changing the initial mass in this way (Fig. \ref{fig:resBhMassGrid}) does not greatly affect the final mass of 
the black holes. Growth ceases at similar values to the fiducial low resolution runs provided that the initial 
mass does not exceed the final mass in the standard case. The star formation rate is similarly independent of initial 
black hole mass (not shown).

\subsection{No star formation}
\label{noSFruns}

We showed previously (Section \ref{IsoGalAGNfdbck}) that in isolated galaxies the growth of the central black hole slows over 
time. This may be interpreted as evidence that the black hole has self-regulated its growth through feedback, however 
we noted that this `self regulation' occurs on a similar timescale as gas is depleted from the galaxy. As a means of modifying 
the timescale over which gas is consumed in the galaxy systems, we perform additional 
simulations in which we artificially turn off star formation. 

Fig. \ref{fig:DGbhMassNOsf} shows black hole mass versus time for models with no star formation. All models yield substantially 
larger final black hole masses and the slowing of black hole growth towards the end of the simulation is less clear. This shows 
that, at least in artificially isolated galaxies without an external gas source, black hole growth is halted due to external 
gas consumption through star formation. Although a galaxy would normally be fed with cooling gas from a 
halo or accretion along a filament, it is possible that the regulation of black hole growth through star formation is important in 
low redshift galaxies for which filamentary structures have been disrupted and/or the gas becomes too diffuse to cool 
efficiently onto the galaxy.

\begin{figure}
\begin{center}
\subfigure{\includegraphics[width=\columnwidth]{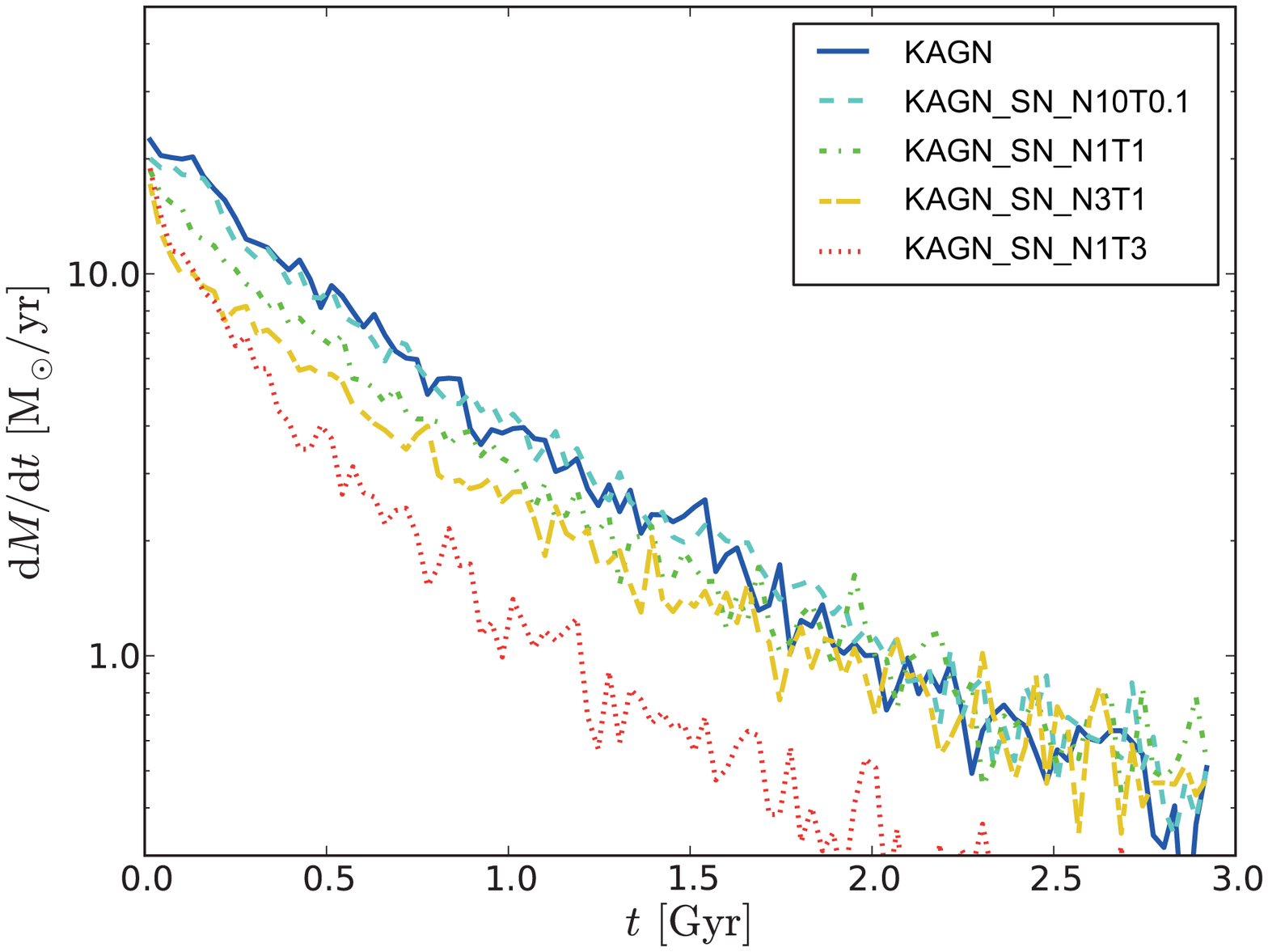}}
\subfigure{\includegraphics[width=\columnwidth]{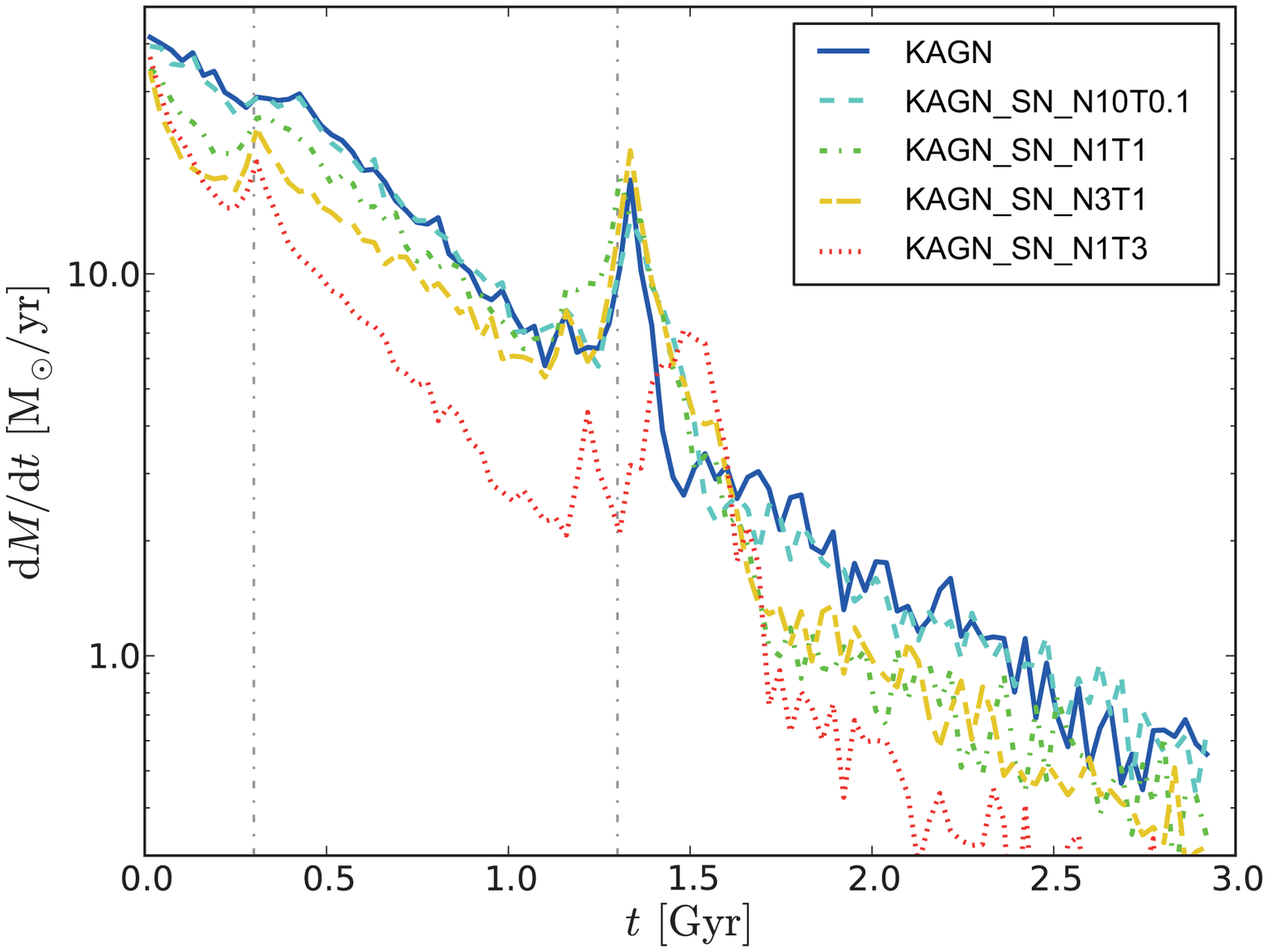}}
\end{center}
\caption{Evolution of the star formation rate with varying supernova feedback strength for low resolution simulations 
employing kernel-weighted, Bondi-type AGN feedback in isolated (top panel) and merging (bottom panel) disc galaxies. 
The low resolution fiducial model (KAGN) is shown for comparison. 
The N1T3 model causes by far the most suppression of star formation.}
\label{fig:SFRsnT}
\end{figure}

\subsection{Supernova feedback parameters}

\label{snTemp}
\begin{table}
\caption{Parameters for supplementary simulations investigating supernova feedback. Columns detail the simulation label, 
supernova feedback efficiency, number of gas particles heated by each star particle and the temperature by which gas is heated 
through a feedback event. Note that KAGN\_SN\_N1T1 is our fiducial choice.}
\begin{center}
\begin{tabular}{p{25mm} c c c c}
\hline
Label & $\epsilon_{\rm SN}$ & SN $N_{\rm heat}$ & $T_{\rm heat}\, [{\rm K}]$ \\
\hline
KAGN\_SN\_N1T1    & 0.37 & 1  & $1\times 10^7$ \\
KAGN\_SN\_N10T0.1 & 0.37 & 10 & $1\times 10^6$ \\
KAGN\_SN\_N3T1    & 1.1  & 3  & $1\times 10^7$ \\
KAGN\_SN\_N1T3    & 1.1  & 1  & $3\times 10^7$ \\
\hline
\end{tabular}
\end{center}
\label{tab:SimparamsSN}
\end{table}

As a probe of the numerical sensitivity of our supernova feedback implementation we also present additional low resolution 
simulations performed with varying mass loading factors and supernova heating temperatures. In the fiducial model, 
each star formed causes one neighbouring gas particle to be 
heated to $10^7\, {\rm K}$, coupling the supernova energy with an efficiency, $\epsilon_{\rm SN}=0.37$. Here we additionally
 simulate this feedback with several different parameter choices (Table \ref{tab:SimparamsSN}), namely heating 10 particles 
to $10^6\, {\rm K}$, thereby keeping the total energy output constant, a case where we deposit all the energy available 
(see Section \ref{SNfeedbackmethod} for a derivation) into 3 particles heated to $10^7\, {\rm K}$; and a maximal case where 
we heat 1 particle to $3\times 10^7\, {\rm K}$. 

Fig. \ref{fig:DGprettySN} shows projected gas density plots for the range of supernova parameters. Increasing the 
strength of the feedback, and particularly the temperature, increases the porosity in the gas disc and redistributes 
more gas into the hot halo. Note that the simulations were performed with the KAGN method in conjunction with supernova 
feedback as this model allows an investigation of the effect on the black hole during a merger, without the AGN terminating 
star formation.

Fig. \ref{fig:SFRsnT} shows the evolution of the star formation rate in isolated (top panel) and merging (bottom panel) 
galaxies. Star formation is barely suppressed in the KAGN\_SN\_N10T0.1 simulation, whilst the fiducial model 
KAGN\_SN\_N1T1 is only marginally less powerful than the KAGN\_SN\_N3T1 variant, despite the latter having three 
times as much energy available per supernova. Finally, the KAGN\_SN\_N1T3 model causes by far the most suppression 
of star formation at all times, even delaying the timing of the starburst. The delay does not have a dynamical origin 
as the black holes are dynamically bound/merged by $t=1.37\, {\rm Gyr}$. It is caused by the longer time it takes 
for the gas to condense out of the hotter and more diffuse halo created by the supernovae.

In summary, the level of star formation suppression due to supernovae is most sensitive to the choice of heating temperature, a finding in agreement with 
the results from our AGN feedback simulations.

\subsection{AGN feedback parameters}
\label{bhStrength}

The findings of our main investigation indicate that the most important aspect of an AGN model, in terms of the effect on 
the host galaxy, are the details of how the feedback energy is deposited. Specifically, the temperature reached by heated 
gas is shown to be critically important. Striking differences were found between the Bondi-based AGN models of 
\cite{DMS2005} and \cite{boothschaye2009} which differ primarily by employing kernel-weighted and strong feedback 
respectively. However, the details of the two methods have additional variations; the values of the feedback coupling 
efficiency ($\epsilon_{\rm f}=0.05$ versus $0.15$) and accretion rate boost, $\alpha=100$ or $\propto \rho^{2}$ 
(see Section \ref{AccretionSect} for details) both change between the models. In order to disentangle the effects of 
these choices we have performed two extra simulations, changing each of the parameters one at a time. 
Table \ref{tab:SimparamsAGNstrength} summarises the extra simulations performed.

\begin{table}
\caption{Additional simulation parameters probing the AGN feedback method. Columns detail the simulation label, 
Bondi accretion rate boost factor, number of feedback-heated particles, feedback distribution type and AGN feedback efficiency.}
\begin{center}
\begin{tabular}{p{18mm} c c c c c}
\hline
Label & Acc. & $\alpha$ & $N_{\rm heat}$ & Type & $\epsilon_{\rm f}$ \\
\hline
KAGN\_SN\_$\alpha$\_05 & Bondi & $100$ & 50 & kernel & $0.05$ \\
SAGN\_SN\_$\alpha$\_05 & Bondi & $100$ & 1 & strong & $0.05$ \\
SAGN\_SN\_$\alpha$\_15 & Bondi & $100$ & 1 & strong & $0.15$ \\
SAGN\_SN\_$\beta$\_15 & Bondi & $\propto n_{\rm H}^2$ & 1 & strong & $0.15$ \\
\hline
\end{tabular}
\end{center}
\label{tab:SimparamsAGNstrength}
\end{table}

\begin{figure}
\begin{center}
\subfigure{\includegraphics[width=\columnwidth]{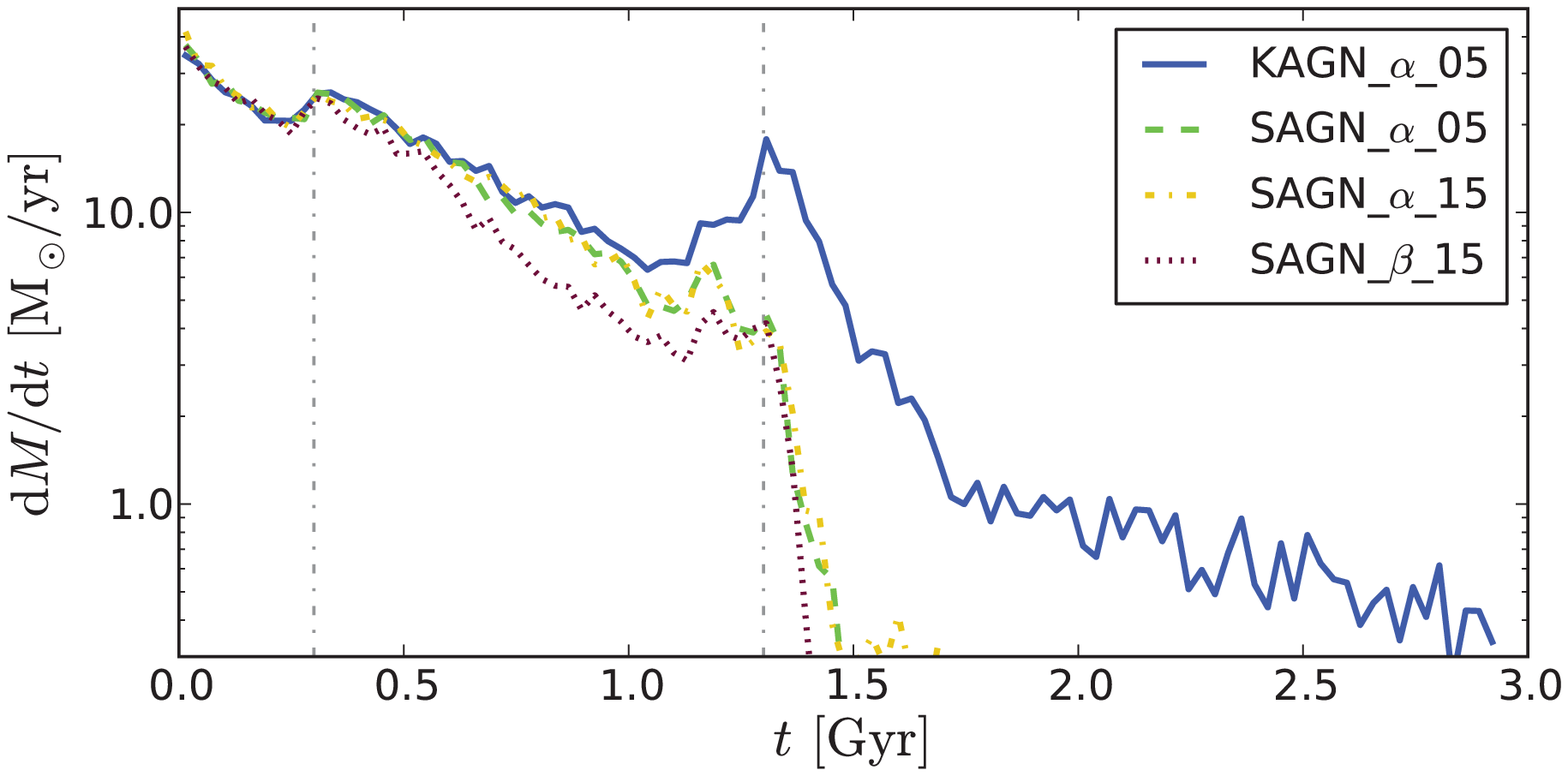}}
\subfigure{\includegraphics[width=\columnwidth]{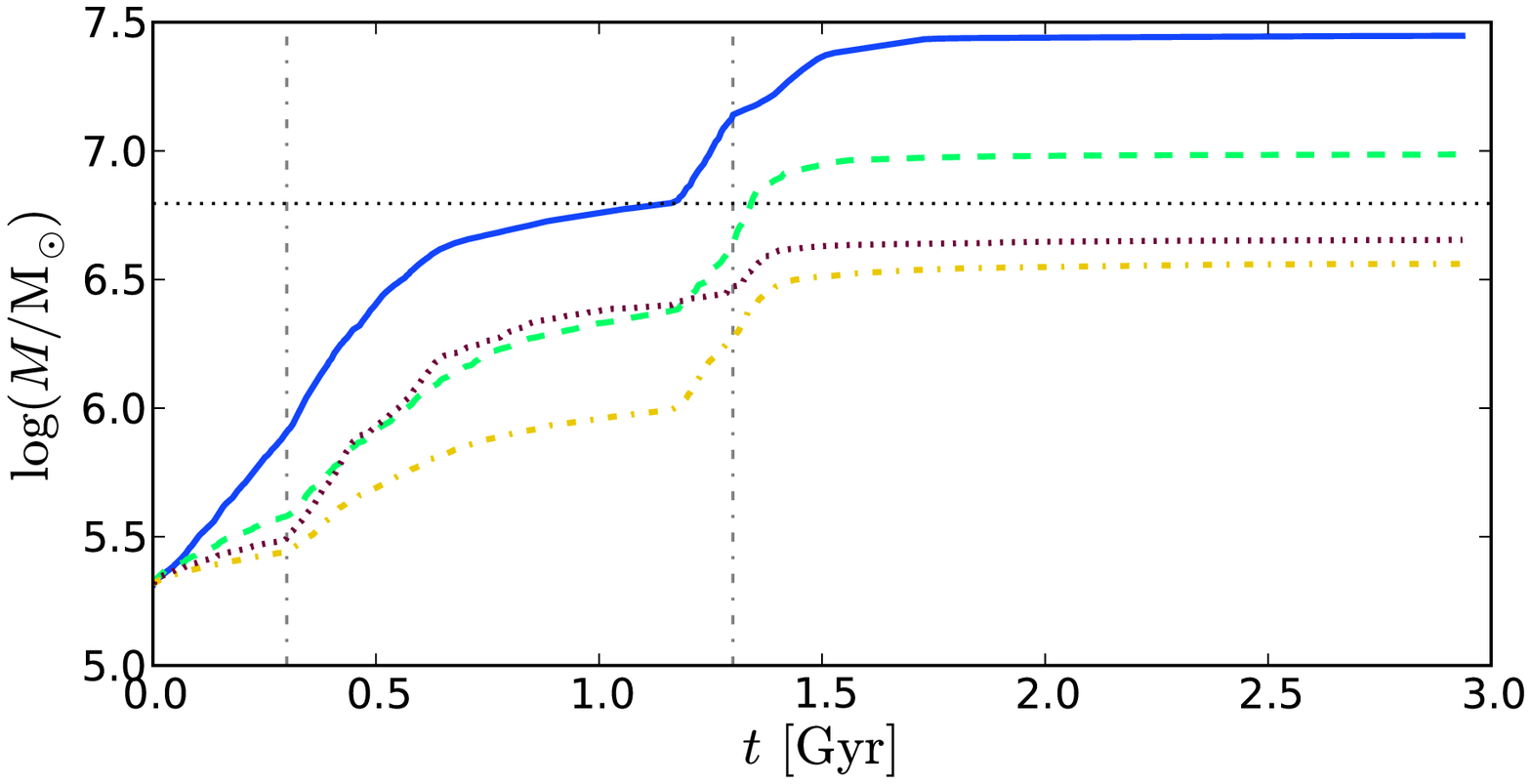}}
\end{center}
\caption{Effect of varying AGN feedback parameters on the star 
formation rate (top panel) and total black hole mass (bottom panel) in low resolution merger simulations. 
The star formation rate through a merger is dominated by the choice of feedback model, while the final black hole 
mass also shows a strong dependence on the efficiency parameter $\epsilon_{\rm f}$.}
\label{fig:BHfdbk}
\end{figure}

In Fig. \ref{fig:BHfdbk}, we compare the star formation rate (top panel) and total black hole mass (bottom panel) between 
the fiducial Bondi models and the additional merger simulations at low resolution. This shows conclusively that the heating method is 
the dominant factor in determining the star formation suppression by AGN feedback in these systems. Changing the 
accretion rate density dependence does have a small effect on the star formation at early times, but the dominant trend of declining 
and terminating star formation still occurs at the same time. Modifying the available energy in feedback by factor of three 
through the $\epsilon_{\rm f}$ parameter has no discernible impact upon the star formation rate (a result similar to that 
found by \citealt{debuhr2011}).

The black hole mass evolution is also sensitive to changes made to the heating efficiency parameters, showing a trend for 
reducing mass with increasing feedback strength as would be expected. These findings show that the black hole feedback 
efficiency can be used to tune the black hole mass without affecting the host galaxy, as long as the feedback method is constant,
in agreement with \cite{boothschaye2009}.

\section[]{Discussionand conclusions}
\label{conclusions}

AGN feedback models are becoming a standard component in both large scale {\it N}-body hydrodynamical simulations and semi-analytic 
modelling, but the full implications of incorporating even the current simplistic models are not yet fully understood. One 
particular uncertainty is how the combination of AGN and supernova feedback processes change the star formation rate in
a galaxy. We performed and compared a series of identical high resolution disc galaxy simulations with a range of sub-grid model 
combinations including a variety of AGN models and supernova feedback strengths. Changing only the feedback models allowed us 
to perform a clean comparison between competing parametrisations, as well as probing the details of the processes involved and how 
the models interact. Beginning with an isolated disc galaxy we investigated how feedback processes change the star formation rate and 
where appropriate the black hole growth. Building on this understanding we simulated the galaxies undergoing major mergers and 
analysed the effects of AGN and supernova feedback on the galaxy through an active phase in its evolution. 
Our main conclusions may be summarised as follows:
\begin{itemize}
\item
In isolated Milky-Way-mass model disc galaxies supernova feedback dominates the star formation suppression, whereas 
AGN feedback has no impact for any of the tested models. This is due to the disparity in scales upon which the processes 
act; AGN activity is (eponymously) confined to the central region whereas supernovae occur throughout the galactic disc.
\item
AGN feedback plays a much larger role in mergers and may completely eliminate the starburst when the heating is strong. 
The most important factor determining the strength of AGN feedback on the host galaxy is 
the temperature to which gas is heated and correspondingly the feedback mass loading. A higher temperature allows (fewer) 
gas particles to escape from high density regions without suffering from significant radiative losses.
\item
Supernova and AGN feedback are largely independent, with only a slightly stronger suppression of the star formation rate when 
acting together than if they acted independently. This finding is in contradiction to the findings of 
\cite{vecchiaschaye2012} and supports the approach taken in semi-analytic modelling, (\citealt{bower2006,croton2006,guo2011}). 
However, our simulations are a restricted set of conditions and ignore complex effects such as external gas accretion.
\item
Numerical resolution strongly affects the black hole mass due to under-resolved feedback and accretion. The problem is more severe 
for the kernel-weighted model where lower resolution simulations result in lower heating temperatures due to the increase 
in mass within the kernel. However, the stellar mass appears to converge less quickly with resolution in strong feedback models.
\item
The black hole mass is sensitive to the choice of both the feedback and accretion models, however the parameters may 
be tuned to obtain desired masses with little effect on the galaxy. The mass is strongly affected by gas consumption 
through star formation in these model systems which neglect an external gas source such as a hot halo or minor mergers. 
Gas fraction and initial black hole mass do not affect the final black hole mass by a large factor.
\item
The level of star formation suppression due to supernovae is (as for AGN feedback) strongly linked to the temperature 
of heated gas with higher temperatures causing a larger reduction. The number of heated particles (mass loading) is 
of secondary importance. However, even the maximally strong supernova feedback model was unable to prevent a 
starburst occurring in the galaxy merger when AGN feedback was weak.
\end{itemize}

The main outcomes of this study will be of use for informing future theoretical works simulating/modelling AGN and SN feedback. Future $N$-body methods incorporating  AGN in galaxy formation simulations should consider that the (thermal) feedback strength is key factor in determining the effect on its host galaxy. Additionally we find that AGN and SN feedback processes are largely independent across models in these systems, supporting the approach taken in semi-analytic modelling. Our findings may indicate that it may be premature to attempt to probe the detailed features of AGN (e.g. durations of feedback episodes, luminosities) for comparison with observations using the current generation of models designed for studying galaxy formation.

While we have gone some way towards disentangling the role of feedback processes across multiple models for the same 
systems, further work is required to make progress with many aspects of our understanding of AGN feedback in galaxies. 
Due to our need to perform multiple high resolution simulations, we were only able to study two identical systems. Future 
work, taking advantage of advances in computational power, may investigate these processes in cosmological zoom 
simulations (\citealt{tormen1997}) of multiple galaxy masses and environments. A possible shortcoming of this work is that 
the IMF employed (Salpeter) is currently in question, with alternatives e.g. Chabrier and Kroupa IMFs arguably giving 
better fits to data. An alternative IMF alters the star formation history of a galaxy 
as well as the energy available from supernovae. However the trends outlined in this paper are unlikely to change substantially 
with an alternative IMF for a given supernova model. .

Despite ongoing work (e.g. \citealt{hobbs2012,debuhr2011,fanidakis2011,planelles2012}), the current generation of AGN 
models such as those incorporated in this study are unable to perform well over the full mass range in cosmological 
objects, from dwarf galaxies to clusters, as a fully developed model should. It is therefore important that the 
development of a flexible, resolution independent AGN feedback model continues.

\vspace{1em}

While this paper was in advanced draft form, two recent papers appeared by \cite{wurster2013agn,wurster2013}. These papers share 
some commonality with this work as they also investigate several AGN models in idealised isolated and merging disc galaxies, 
including the \cite{power2011} disc accretion model. The papers are complementary to our own as they choose to focus primarily 
on the details of the AGN models themselves, rather than their interaction with the host galaxies. Despite the differing focus, some results may be compared with our own. As found in our simulations for all models, \cite{wurster2013} note that lower resolution DAGN simulations yield more massive black holes. Across the models considered \cite{wurster2013agn} also find that although the final BH masses may be tuned to obtain similar values, the pattern of growth differs between models. Furthermore, a stronger star formation 
suppression is found for strong AGN feedback than kernel-weighted feedback (in their corresponding models), in agreement with 
our own findings.

\section*{Acknowledgments}

The simulations used in this paper were performed on the ICC Cosmology Machine, which is part of the DiRAC Facility jointly 
funded by STFC, Large Facilities Capital Fund of BIS and Durham University. 
RDAN acknowledges the support of an STFC studentship. 
Rendered images were processed using SPLASH (http://users.monash.edu.au/$\sim$dprice/splash/). 

\bibliographystyle{mn2eFix}
\bibliography{AGN_SN}

\appendix

\section[]{Creating equilibrium disc galaxies}
\label{ICappendix}
In creating our model galaxies we largely follow \cite{SDH2005modelling}; however our approach differs in a few key areas such as our 
treatment of adiabatic contraction of the DM halo (see Section \ref{Hcontract}) and the ISM model used (Section 
\ref{radCoolingISMSect}). The model galaxies created here will consist of an encompassing DM halo surrounding a disc made up of stars 
and gas particles with an optional stellar bulge.

\subsection{Collisionless particle positions}
When defining the positions of the collisionless components we begin with a so-called `{\it glass}', a volume of random and 
homogeneously distributed particles. We then modify the glass by adding additional constraints on the distribution of the particles 
to fit desired profiles. In this way any undefined aspect of the particle distribution is inherently random and we do not rely solely 
on pseudo-random number generators. Should the glass in use contain insufficient particles we copy and tile the existing glass until 
we obtain a cube large enough.
\subsubsection{Dark matter halo \& the stellar bulge}

For the DM component we assume that it follows the Hernquist mass profile \cite{hernquist1990}
\begin{equation}
\rho_{\rm h}(r) = \frac{M_{\rm h}}{2\pi} \frac{a}{r(r+a)^3},
\end{equation}
where $M_{\rm h}$ is the total halo mass ($M_{\rm h}=f_{\rm h}M$ where $M$ is the total galaxy mass and $f_{\rm h}$ is the halo mass fraction) and $a$ is a constant scale length 
which affects the central shape of the profile. The corresponding enclosed mass profile is then
\begin{equation}
\label{DMmass}
M_{\rm h}(<r) = M_{\rm h} \frac{r^2}{(r+a)^2}.
\end{equation}
We choose this profile as it replicates the inner slope of the cosmologically motivated NFW profile (\citealt{NFW1996}) whilst having 
a mass profile which converges at large radii. We utilise the following conversion between $a$ and the NFW concentration parameter, 
$c$, under the assumption that the two profiles contain the same mass within $r_{200}$ (the radius at which the density of the system 
drops below $200\times$ the critical density required to close the universe at that redshift)
\begin{equation}
a=r_{\rm s}\sqrt{2[\ln(1+c)-c/(1+c)]},
\end{equation}
where $r_{\rm s}=r_{200}/c$ is the scale length of the NFW halo. We then distribute the DM particles simply by radially sliding the particles to the 
position found by interpolating Equation \ref{DMmass} at $M(<r)=(N+1)m_{\rm DM}$, where $N$ is the number of particles already placed and 
$m_{\rm DM}$ is the mass of a DM particle. In this way the random angular position of the particle is maintained.

For the stellar bulge we also follow a Hernquist profile with the modification that a different (larger, \citealt{NFW1996}) 
scale length $b$ is employed instead of $a$ and the particle mass is now $m_{\rm star}$
\begin{equation}
\label{bulgemass}
M_{\rm b}(<r) = M_{\rm b} \frac{r^2}{(r+b)^2}.
\end{equation}
Both $a$ and $b$ are free parameters of our model.

\subsubsection{Stellar disc}
\label{SDpos}
For the stellar disc we define that its surface density is described by an axisymmetric exponential disc with scale length $h$
\begin{equation}
\label{stelsurf}
\Sigma_{\star}(R)=\frac{M_{\star}}{2\pi h^2} e^{(-R/h)},
\end{equation}
where $M_{\star}(=(1-f_{\rm g})M_{\rm d})$ is the mass of the stellar disc and $f_{\rm g}$ is the disc gas mass fraction 
($M_{\rm d}=M_{\star}+M_{\rm g}$). From this we deduce the enclosed mass profile and perform the same procedure as 
followed for the halo when setting the particle radii, sliding in $R$
\begin{equation}
M_{\star}(<R)=M_{\star}\frac{(R+h)}{h}e^{(-R/h)}.
\end{equation}
The scale length, $h$, is set for both gaseous and stellar components by relating it to the angular momentum of the disc. We calculate 
the angular momentum by assuming that the baryonic matter which now makes up the disc was initially distributed identically to the 
surrounding DM halo forming a `total' halo and then applying conservation of angular momentum. In this case the angular momentum of 
the disc is equal to a fraction of the `total' halo angular momentum, i.e.
\begin{equation}
J_{\rm d}=j_{\rm d}J,
\end{equation}
 where we take the approach of \cite{springwhite1999} in assuming there is no angular momentum transport between the different galaxy 
components, under this assumption the angular momentum fraction, $(j_{\rm d}$, is equal to the mass fraction of the disc. The spin of 
a halo is often described by the dimensionless spin parameter, $\lambda$
\begin{equation}
\label{lambd}
\lambda=\frac{J{|E|}^{1/2}}{GM^{5/2}},
\end{equation}
where $E$ is the total energy of the halo. For an NFW halo the kinetic energy is found to be, under the assumption that all particles 
move on circular orbits at the circular velocity (\citealt{springwhite1999})
\begin{equation}
E_{\rm kin}=\frac{GM^2}{2r_{200}}f_{\rm c},
\end{equation}
where
\begin{equation}
f_{\rm c}=\frac{ c \left[ 1 - \frac{1}{{(1+c)}^2} - \frac{2\ln(1+c)}{1+c} \right] }{ 2\left[ \ln(1+c) - \frac{c}{1+c} \right]^2 }.
\end{equation}
Although this relationship is strictly true for NFW profiles we assume it also holds for the Hernquist profile due to the similarity 
of the profiles in the central regions. Then applying the Virial theorem so $E=-E_{\rm kin}$ and by substituting into Equation \ref{lambd} 
we obtain an expression for the halo angular momentum
\begin{equation}
J=\lambda G^{1/2} M_{200}^{3/2} r_{200}^{1/2} \left(\frac{2}{f_{\rm c}}\right)^2,
\end{equation}
We may then find a $h$ value for which the numerically calculated $J_{\rm d}$ value agrees with the predicted value (in practice DM halo 
contraction must be simultaneously considered, see Section \ref{Hcontract}). The angular momentum of any disc with known rotation 
curve and surface density may be found with (\citealt{MMW})
\begin{equation}
J_{\rm d} = M_{\rm d}\int_0^\infty V_{\rm c}(R)R\times2\pi R\Sigma(R)\,\,{\rm d}R,
\end{equation}
where $V_{\rm c}$ is the circular velocity. For our model we substitute in Equation \ref{stelsurf} to find
\begin{equation}
J_{\rm d} = M_{\rm d}\int_0^\infty V_{\rm c}(R) \left(\frac{R}{h}\right)^2\exp\left(-\frac{R}{h}\right){\rm d}R
\end{equation}
where $M_{\rm d}=$ is the total disc mass, and the circular velocity in our system is given by (e.g. \citealt{binneytremaine1987} pg. 77),
\begin{multline}
V_{\rm c}^2(R)=\frac{ G [ M_{\rm h}(<R)+M_{\rm b}(<R) ] }{ R }\\
 + \frac{ 2GM_{\rm d} }{ h }y^2 [I_0(y)K_0(y)-I_1(y)K_1(y)],
\end{multline}
where $y=R/(2h)$ and $I_n$, $K_n$ are Bessel functions.

For the vertical distribution we employ the model of an isothermal sheet with constant scale height $z_0$. From this we once again find the enclosed mass profile which can be rearranged to give the vertical positions
\begin{equation}
\Sigma_{\star}(z)=\frac{M_{\star}}{2z_0}{\rm sech}^2\left(\frac{z}{2z_0}\right),
\end{equation}
\begin{equation}
M_{\star}(<z)=M_{\star}{\rm tanh}\left(\frac{z}{2z_0}\right).
\end{equation}
Here we follow \cite{SDH2005modelling} in setting $z_0=0.2h$. For completeness, the 3D density and mass distribution of the stellar disc are described by
\begin{equation}
\rho_{\star}(R,z)=\frac{M_{\star}}{4\pi z_0h^2}{\rm sech}^2\left(\frac{z}{2z_0}\right) e^{(-R/h)},
\end{equation}
\begin{equation}
\label{sdmenc}
M_{\star}(<R,<z)=\frac{M_{\star}}{h}\tanh \left( \frac{z}{2z_0}\right)  \left( h- e^{(-R/h)}(R+h) \right),
\end{equation}

We note here a numerical consideration that the $R$, $z$ coordinates of a particle are independent, therefore any artificial correlation must be carefully avoided.

\begin{figure}
\begin{center}
\subfigure{\includegraphics[width=\columnwidth]{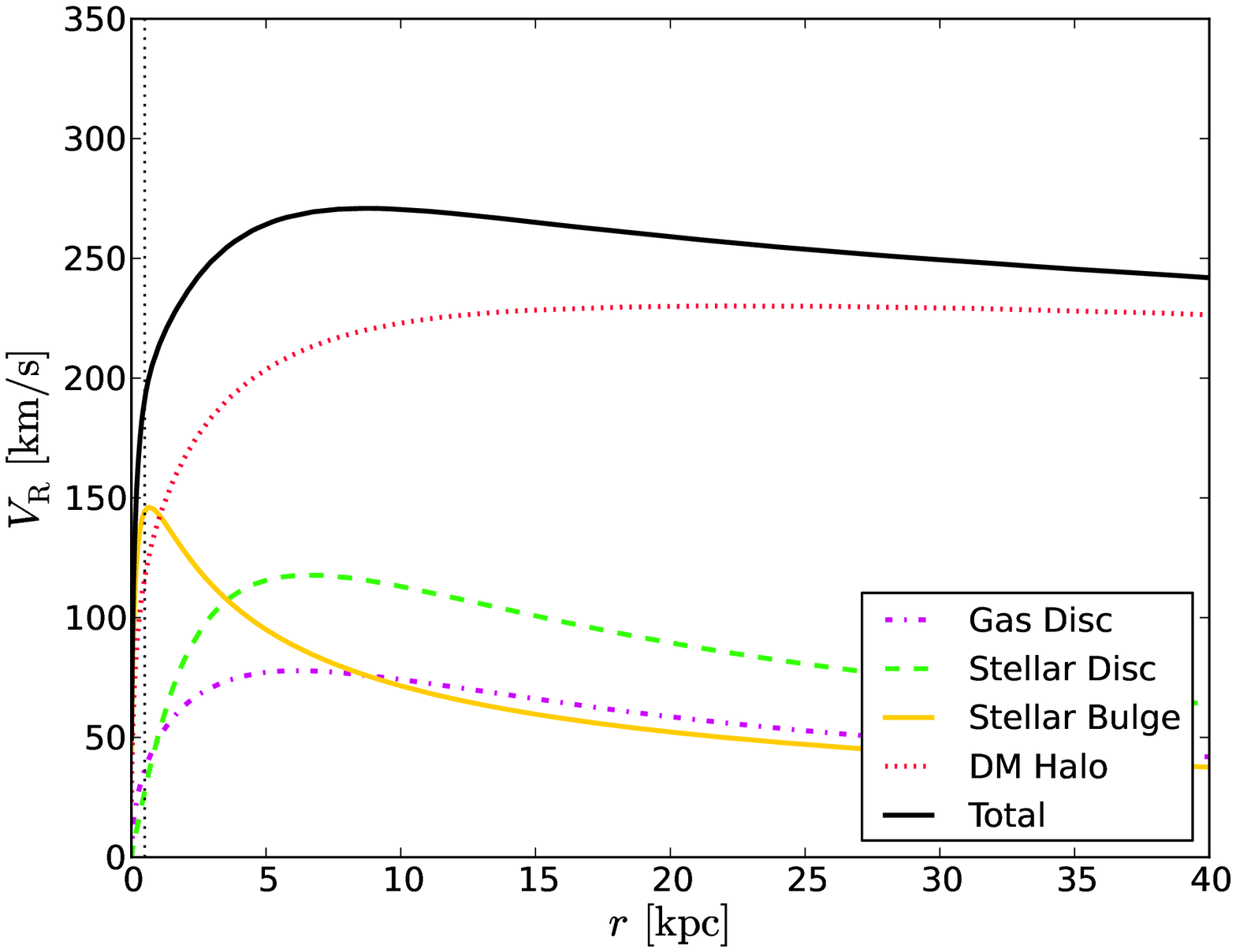}}
\subfigure{\includegraphics[width=\columnwidth]{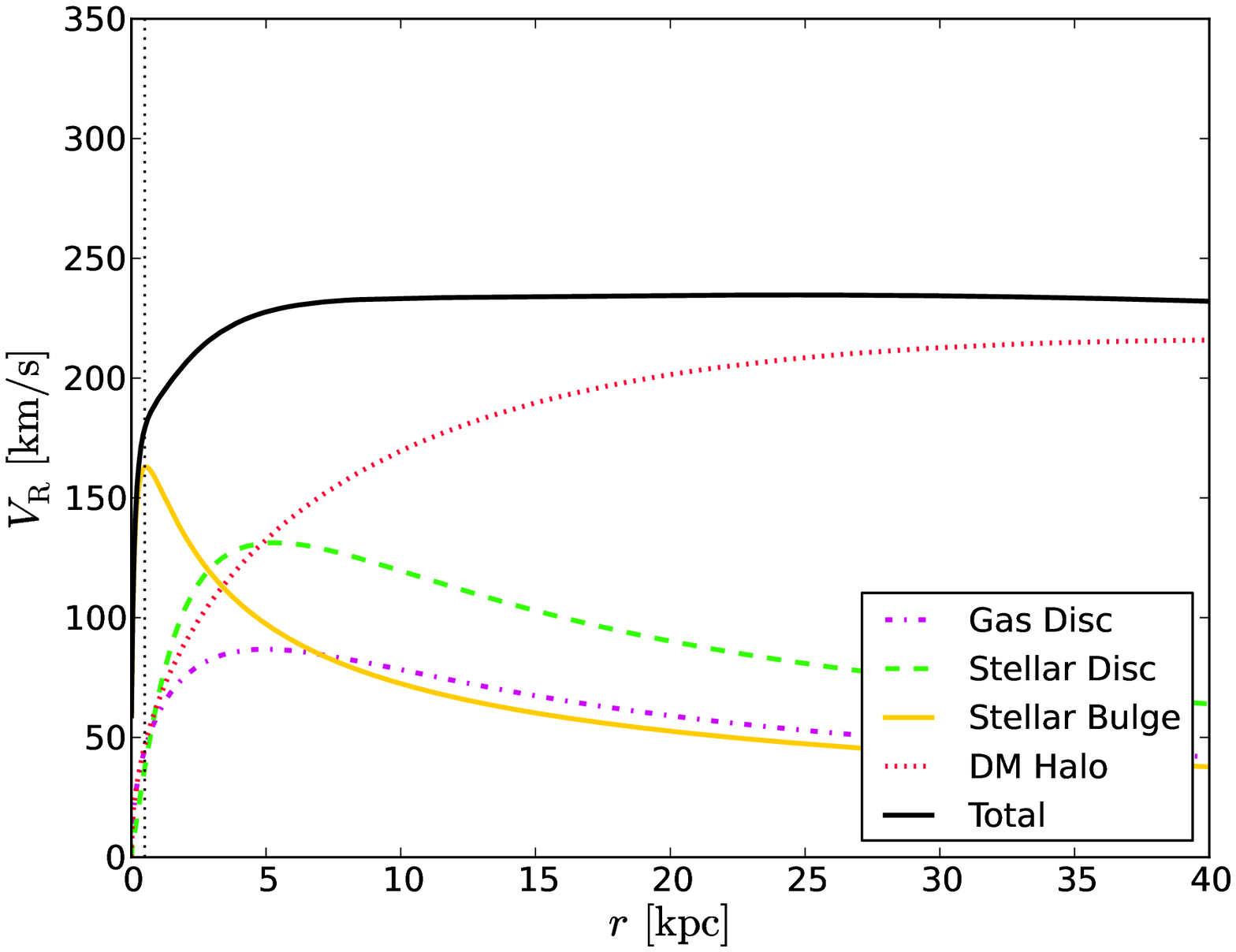}}
\end{center}
\caption{Rotation curves as derived with (top panel) and without (bottom panel) adiabatic contraction of the DM halo. Also shown as a 
vertical dotted line is $2.8\,\varepsilon_{\rm soft}$ where $\varepsilon_{\rm soft}$ is the Plummer equivalent softening length. Note that the two cases have differing disc scale lengths, $h=3.45\, {\rm kpc}$ and $2.78\, {\rm kpc}$ respectively, this is due to the dependence of the scale length on the DM halo circular velocity and therefore its contraction.}
\label{fig:rotCurves}
\end{figure}

\subsubsection{Halo contraction}
\label{Hcontract}
As an optional feature we include the ability to approximate the DM halo contraction resulting from the formation of a galactic disc. Following the approach of \cite{blumenthal1985} (also \citealt{MMW}) we assume that the matter which makes up the galaxy, both dark and otherwise, was initially distributed following a Hernquist profile (as employed in our calculations of the disc angular momentum) but as the baryonic matter began to cool and collapse the DM distribution responded by contracting adiabatically. We also assume that the halo remains spherically symmetric under the contraction, under these assumptions the angular momentum of individual DM particles is conserved. Considering a single particle at initial radius,$r_{\rm i}$, moving to a new position with radius, $r_{\rm f}$, we may write
\begin{equation}
\label{contract}
GM_{\rm f}(<r_{\rm f})=GM_{\rm i}(<r_{\rm i}),
\end{equation}
where $M_{\rm i}$ is the total initial mass enclosed within $r_{\rm i}$ as given by the Hernquist profile and $M_{\rm f}$ is the total final mass enclosed within $r_{\rm f}$. The final mass $M_{\rm f}$ is equal to the sum of the (now contracted) DM component of the mass initially within $r_{\rm i}$ ($=f_{\rm h} M_{\rm i}(<r_{\rm i})$) and the baryonic matter within $r_{\rm f}$
\begin{equation}
M_{\rm f}(<r_{\rm f})=M_{\rm d}(<r_{\rm f}) + M_{\rm b}(<r_{\rm f})+f_{\rm h}M_{\rm i}(<r_{\rm i}),
\end{equation}
where we analytically know $M_{\rm b}(<r)$ from Equation \ref{bulgemass} and $M_{\rm d}(<r)$ is found similarly to Equation \ref{sdmenc} whilst setting $R=z=r$
\begin{equation}
M_{\rm d}(<r)=\frac{M_{\rm d}}{h}\tanh \left( \frac{r}{2z_0}\right)  \left( h- e^{(-r/h)}(r+h) \right).
\end{equation}
By iterating we then find a value of $r_{\rm f}$ which satisfies Equation \ref{contract}. The problem is further complicated because the determinations of the disc scale length, $h$, and the halo contraction are interdependent; we must therefore iterate over the two calculations from a first guess until the solutions converge. See Fig. \ref{fig:rotCurves} for an illustration of the effects of contraction on rotation curves.

\subsubsection{Gaseous disc}
\label{vertgas}
The radial dependence of the gas distribution is set as for the stellar component in Section \ref{SDpos}, and follows
\begin{equation}
\Sigma_{\rm g}(R)=\frac{M_{\rm g}}{2\pi h^2} e^{(-R/h)}.
\end{equation}

To ensure the gas is initially placed in a stable structure we assume hydrostatic equilibrium
\begin{equation}
\frac{\partial P}{\partial z} = - \rho_{\rm g} \frac{\partial \Phi}{\partial z},
\end{equation}
where $\Phi$ is the total gravitational potential. Substituting in our EoS (Equation \ref{EoS}) we obtain
\begin{equation}
\begin{split}
\frac{\partial P}{\partial z} & = A \gamma \rho_{\rm g}^{\gamma-1} \frac{\partial \rho_{\rm g}}{\partial z}, \\
& =  - \rho_{\rm g} \frac{\partial \Phi}{\partial z}.
\end{split}
\end{equation}
Rearranging we see
\begin{equation}
\rho_{\rm g}^{\gamma-2} \frac{{\rm d} \rho_{\rm g}}{{\rm d} z} = -\frac{1}{A\gamma} \frac{{\rm d} \Phi}{{\rm d} z},
\end{equation}
separating variables then gives us the integrals
\begin{equation}
\int_{\rho_0}^{\rho_{\rm g}} \rho_{\rm g}^{\gamma-2} {\rm d}\rho_{\rm g} = -\frac{1}{A\gamma} \int_{\Phi_0}^\Phi {\rm d}\Phi,
\end{equation}
where $\rho_0$,$\Phi_0$ are the respective mid-plane values. Performing the integrations and rearranging gives
\begin{equation}
\label{Grho}
\rho_{\rm g}(z,R) = \left[   \rho_0^{\gamma-1}(R) - \left(  \frac{\gamma-1}{A\gamma} \right) [\Phi(z,R) - \Phi_0(R)]     \right]^{\frac{1}{\gamma-1}}.
\end{equation}
The mid-plane potential $\Phi_0$ is known numerically; however we must determine the mid-plane density, $\rho_0$, by iteration demanding that the 3D density agrees with the surface density i.e. that it fulfills the constraint, 
\begin{equation}
\Sigma_{\rm g}(R) =\int_{-\infty}^\infty \rho_{\rm g}(R,z, \Phi) {\rm d}z.
\end{equation}
At a given radius we may then obtain a profile for $M_g(<z)$ by numerically integrating Equation \ref{Grho}, we then interpolate this profile when vertically positioning the gas particles. The known density distribution is then used with Equation \ref{EoS} and the ideal gas law to find the temperature of each particle.

At this stage we discuss an important numerical consideration in this method, many of the preceding derivations have required the use of spatial derivatives and we have begun to introduce the use of the potential, $\Phi$, for both of these purposes we create a fine logarithmic mesh in $R$, $z$. It is at these mesh points which we calculate the gravitational potential due to the mass distribution using a parallelised `Tree' method and define the density profiles. The mesh is also employed extensively when calculating the velocity distributions.

\subsection{Particle velocities}

\begin{figure}
\begin{center}
\includegraphics[width=\columnwidth]{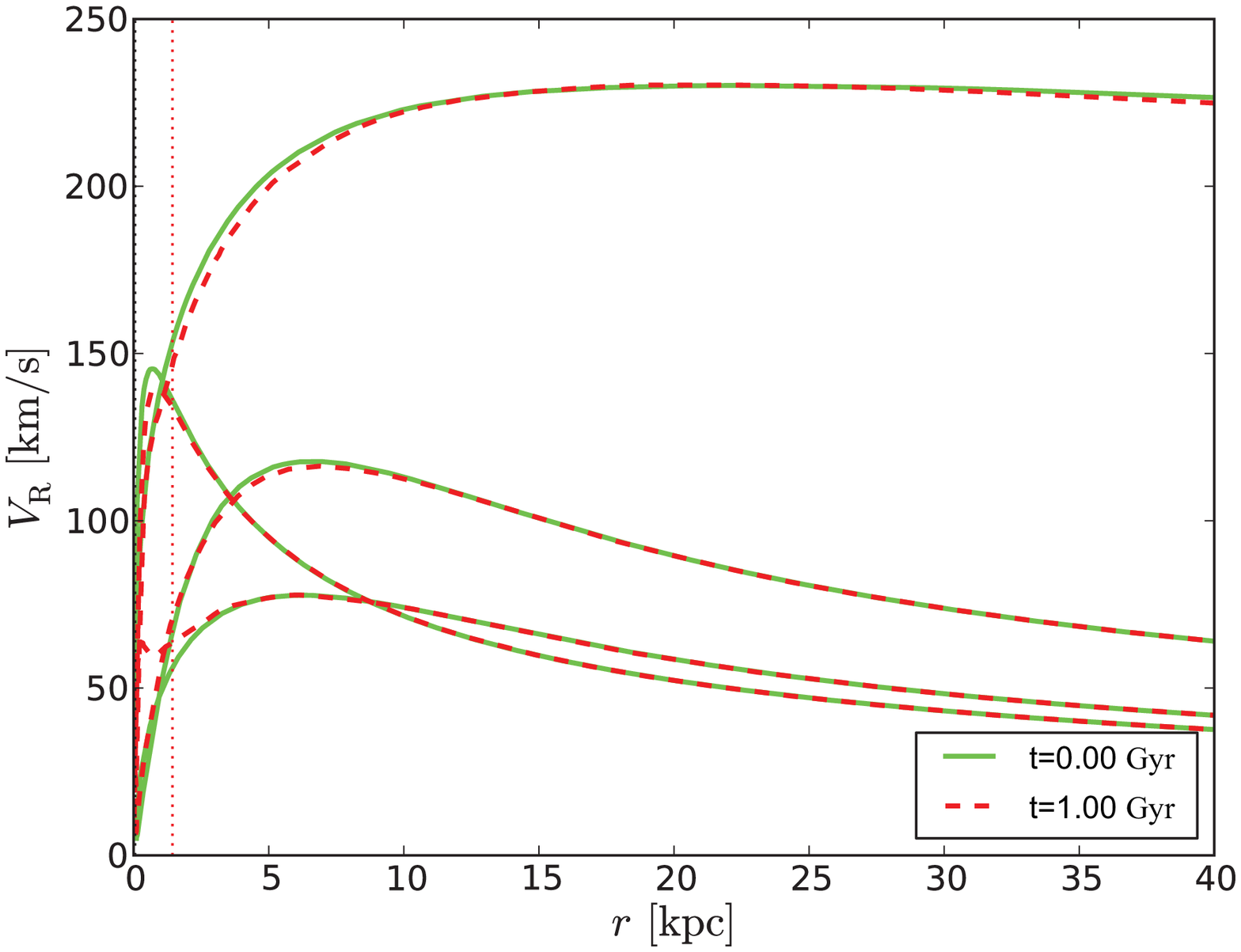}
\end{center}
\caption{Evolution of galaxy rotation curves for an adiabatically contracted isolated model galaxy with $7\times 10^6$ particles. 
Rotation curves are shown at $t=0.0$ and $3.0\, {\rm Gyr}$ in a gravity and hydrodynamics only simulation for (top to bottom) 
dark matter, stellar disc, gas disc and stellar bulge. Also shown as a vertical dotted line is the estimated two-body relaxation 
radius at $1.0\, {\rm Gyr}$. A small level of contraction in the central region is seen, however for the most part the galaxies 
are extremely stable.}
\label{fig:rotCurvesEvo}
\end{figure}

\subsubsection{Collisionless particle velocities}
When considering the velocities of the particles representing collisionless components we follow \cite{hernquist1993} (see also \citealt{binneytremaine1987}) and calculate the distribution by considering the collisionless Boltzmann equation (CBE). In deriving our distribution we assume that it only depends on the energy, $E$, and the $z$-component of the angular momentum $L_{\rm Z}$ as well as assuming that the velocities are isotropic. In such a case the mixed second order moments vanish as do the radial and vertical first order moments ($\langle v_{\rm R}v_{\rm Z}\rangle=\langle v_{\rm R}v_\phi\rangle=\langle v_{\rm Z}v_\phi\rangle=0$, $\langle v_{\rm R}\rangle=\langle v_{\rm Z}\rangle=0$). The non-vanishing second order moments are then obtained from the Jeans equations as
\begin{equation}
\langle v_{\rm Z}^2\rangle=\langle v_{\rm R}^2\rangle=\frac{1}{\rho} \int_{\rm Z}^\infty\rho(z',R)\frac{\partial \Phi}{\partial z} {\rm d} z',
\end{equation}
\begin{equation}
\label{dispPhi}
\langle v_\phi^2\rangle=\langle v_{\rm R}^2\rangle+\frac{R}{\rho} \frac{\partial (\rho \langle v_{\rm R}^2\rangle)}{\partial R} + v_{\rm c}^2,
\end{equation}
where $\rho$ is the density of the currently considered mass component, $\Phi$ is the gravitational potential from all components and $v_{\rm c}$ is the circular velocity
\begin{equation}
v_{\rm c}^2= R\frac{\partial \Phi}{\partial R}.
\end{equation}
We then require a suitable function to approximate the velocity distribution at any point, for this purpose we choose a triaxial Gaussian distribution with its axes coincident with the axisymmetric coordinate system
\begin{equation}
F(v_X, R, z) = \frac{1}{\sqrt{(2\pi \sigma_X^2)}} \exp{\left(-\frac{ (v_X-\mu)^2}{2\sigma_X^2}\right)},
\end{equation}
where $X(=R,z,\phi)$, $\sigma_X^2(=\sigma_X^2(R,z))$ is the velocity variance along a general axis, $X$, and $\mu(=\mu(R,z)=\langle v_X\rangle)$ is the mean streaming velocity. The velocity dispersions and variances are related by
\begin{equation}
\label{disps}
\begin{split}
\sigma_{\rm R}^2 &= \langle v_{\rm R}^2\rangle \\
\sigma_{\rm Z}^2 &= \langle v_{\rm Z}^2\rangle \\
\sigma_\phi^2 &= \langle v_\phi^2\rangle - {\langle v_\phi\rangle}^2
\end{split}
\end{equation}
where we have used $\langle v_{\rm R}\rangle=\langle v_{\rm Z}\rangle=0$.

For the stellar bulge we set the streaming velocity, $\mu$, to zero as in \cite{SDH2005modelling} however this is not true for the DM halo. For the halo we set the mean rotational velocity at any point to be a fixed fraction of the circular velocity (with a fiducial value of $\mu=0.01 v_{\rm c}$), note however that this is of small importance when compared to the radial motions of the DM particles and it is the velocity dispersion which contributes the majority of support to the halo.

Once more we take the approach of \cite{SDH2005modelling} and continue to employ the Gaussian approximation when describing the stellar disc with the assumption that
\begin{equation}
\langle v_{\rm R}^2\rangle=f_{\rm R} \langle v_{\rm Z}^2\rangle,
\end{equation}
where it is assumed that $f_{\rm R}=1$, this is justified as being well supported by observational evidence. In the case of the disc, unlike the halo and bulge, it is the rotational motions of the particles which is most important for providing support to the structure. When considering this mean azimuthal streaming of the disc we invoke the epicyclic approximation
\begin{equation}
\sigma_\phi^2=\sigma_{\rm R}^2\frac{\kappa^2}{4\Omega},
\end{equation}
where the epicyclic frequency, $\kappa$ is defined as
\begin{equation}
\kappa^2=\frac{3}{R} \frac{\partial \Phi}{\partial R} + \frac{\partial^2 \Phi}{\partial R^2},
\end{equation}
and
\begin{equation}
\Omega=\frac{1}{R^2} \frac{\partial \Phi}{\partial R}.
\end{equation}
We then define
\begin{equation}
\eta^2=\frac{4\Omega^2}{\kappa^2}=\frac{4}{R} \frac{\partial \Phi}{\partial R} \left(\frac{3}{R} \frac{\partial \Phi}{\partial R} + \frac{\partial^2 \Phi}{\partial R^2}\right)^{-1}.
\end{equation}
to obtain the simplified relation, $\sigma_\phi^2=\sigma_{\rm R}^2/\eta^2$. Equations \ref{dispPhi} and \ref{disps} then allow us to write the mean streaming velocity
\begin{equation}
\langle v_\phi \rangle = \sqrt{ \langle v_\phi^2 \rangle - \frac{\sigma_{\rm R}^2}{\eta^2} }.
\end{equation}

Numerically, the velocity distributions are defined for the fixed spatial mesh described briefly in Section \ref{vertgas} once the matter distributions have been finalised. This is so that the full gravitational potential is then known in regularly distributed positions and then may be integrated or differentiated as needed. When finding the velocity of any given particle, bilinear interpolation is used to find the approximate $\sigma$ (and where appropriate $\mu$) at the particle's position, the Gaussian distribution is then sampled using the so-called {\it accept-reject} method.

\subsubsection{Gaseous disc velocities}
The velocity structure of the gaseous disc is much simpler because the majority of its support is provided by its pressure, it is therefore 
assumed to undergo solid body rotation i.e. its velocity depends only on $R$ and $v_{\rm R}=v_{\rm R}=0$. Assuming balance between the gravitational pull and the centrifugal and pressure support on the disc we have
\begin{equation}
\frac{\partial \Phi}{\partial R}=\frac{v_{\phi,{\rm gas}}^2}{R}  - \frac{1}{\rho_{\rm g}} \frac{\partial P}{\partial R},
\end{equation}
where $P$ and $\rho_{\rm g}$ are the gas pressure and densities respectively. We then find the particle velocities to be
\begin{equation}
v_{\phi,{\rm gas}}=\sqrt{R \left(  \frac{\partial \Phi}{\partial R} + \frac{1}{\rho_{\rm g}} \frac{\partial P}{\partial R}  \right)}.
\end{equation}

The model galaxy is now fully specified and from the method and assumptions we have utilised we expect it to show very little evolution over multiple dynamical times, as shown in Fig. \ref{fig:rotCurvesEvo}.

\label{lastpage}

\end{document}